\documentclass[journal,comsoc]{IEEEtran}
\usepackage[T1]{fontenc}
\usepackage[english]{babel}

\usepackage{cite}
\usepackage{color}
\ifCLASSINFOpdf
   \usepackage[pdftex]{graphicx}

\else
  
  \usepackage[dvips]{graphicx}
  
\fi

\usepackage{amsmath}
\usepackage[cmintegrals]{newtxmath}
\usepackage{bm}
\usepackage{graphicx}

\usepackage{algorithmic}

\usepackage{array}

\usepackage{lipsum}  

\ifCLASSOPTIONcompsoc
  \usepackage[caption=false,font=normalsize,labelfont=sf,textfont=sf]{subfig}
\else
  \usepackage[caption=false,font=footnotesize]{subfig}
\fi

\usepackage{fixltx2e}
\usepackage{dblfloatfix}

\ifCLASSOPTIONcaptionsoff
  \usepackage[nomarkers]{endfloat}
 \let\MYoriglatexcaption\caption
 \renewcommand{\caption}[2][\relax]{\MYoriglatexcaption[#2]{#2}}
\fi

\usepackage{url}

\newcommand{\module}[1]{\left|#1\right|}
\newcommand{\EX}[1]{\mathbb{E} \left[#1\right]}
\newcommand{\HE}{\textbf{H}_{\text{E}}}
\newcommand{\HB}{\textbf{H}_{\text{B}}}
\newcommand{\vb}{\textbf{v}_{\text{B}}}
\newcommand{\ve}{\textbf{v}_{\text{E}}}
\newcommand{\spread}{\textbf{S}}

\newcommand{\w}{\textbf{w}}

\newcommand{\vect}[1]{\boldsymbol{\mathrm{#1}}}
\newcommand{\mat}[1]{\boldsymbol{\mathrm{#1}}}

\newcommand{\C}{\mathbb{C}}
\usepackage{xfrac}
\newcommand\independent{\protect\mathpalette{\protect\independenT}{\perp}}
\def\independenT#1#2{\mathrel{\rlap{$#1#2$}\mkern2mu{#1#2}}}

\usepackage{hyperref}

\usepackage{tabularx}
\newcolumntype{L}{>{\raggedright\arraybackslash}X}
\begin{document}
	
\title{Physical Layer Security in a SISO Communication using Frequency-Domain Time-Reversal OFDM Precoding and Artificial Noise Injection}

\author{\IEEEauthorblockN{Sidney J. Golstein\IEEEauthorrefmark{1}\IEEEauthorrefmark{2},
		Fran\c cois Rottenberg\IEEEauthorrefmark{1}\IEEEauthorrefmark{3},
		Fran\c cois Horlin\IEEEauthorrefmark{1}, 
		Philippe De Doncker\IEEEauthorrefmark{1}, and
		Julien Sarrazin\IEEEauthorrefmark{2}} \\
\IEEEauthorblockA{\IEEEauthorrefmark{1}Wireless Communication Group,
	Universit\'{e}  Libre de Bruxelles, 1050 Brussels, Belgium} \\
\IEEEauthorblockA{\IEEEauthorrefmark{2}Sorbonne Université, CNRS, Laboratoire de Génie Electrique et Electronique de Paris, 75252, Paris, France \\
	Université Paris-Saclay, CentraleSupélec, CNRS, Laboratoire de Génie Electrique et Electronique de Paris, 91192, Gif-sur-Yvette, France} \\
\IEEEauthorblockA{\IEEEauthorrefmark{3} ICTEAM, Universit\'{e} catholique de Louvain, 1348 Louvain-la-Neuve, Belgium} \\
\IEEEauthorblockA{\small{\{sigolste,fhorlin,philippe.dedoncker\}@ulb.ac.be}} \\
\IEEEauthorblockA{\small{francois.rottenberg@uclouvain.be}}\\
\IEEEauthorblockA{\small{julien.sarrazin@sorbonne-universite.fr}}
\thanks{This work was supported by the ANR GEOHYPE project, grant ANR-16-CE25-0003 of the French Agence Nationale de la Recherche and was also carried out in the framework of COST Action CA15104 IRACON.}}

\markboth{Paper-TW-Nov-20-1594}
{Shell \MakeLowercase{\textit{et al.}}: Bare Demo of IEEEtran.cls for IEEE Communications Society Journals}

\maketitle

\begin{abstract}
A frequency domain (FD) time-reversal (TR) precoder is proposed to perform physical layer security (PLS) in single-input single-output (SISO) systems using orthogonal frequency-division multiplexing (OFDM) and artificial noise (AN) signal injection. The AN signal does not corrupt the data transmission to the legitimate receiver but degrades the decoding performance of the eavesdropper. This scheme guarantees the secrecy of a communication towards a legitimate user when the transmitter knows the instantaneous channel state information (CSI) of the legitimate link thanks to the channel reciprocity in time division duplex (TDD) systems, but does not know the instantaneous CSI of a potential eavesdropper. Three optimal decoding structures at the eavesdropper are considered in a fast fading (FF) environment depending on the handshake procedure between Alice and Bob. Closed-form approximations of the AN energy to inject in order to maximize the SR of the communication are derived. In addition, the required conditions at the legitimate receiver's end to guarantee a given SR are determined when Eve's signal-to-noise ratio (SNR) is infinite. Furthermore, a waterfilling power allocation strategy is presented to further enhance the secrecy of the scheme. Simulation results are presented to demonstrate the security performance of the proposed secure system.
\end{abstract}


\IEEEpeerreviewmaketitle

\section{Introduction}

\IEEEPARstart{I}{nternet}-based services have become ubiquitous in daily life. Wireless communication has become the dominant access for most of these services but it is intrisically unsecure due to its unbounded nature. Therefore, several issues have emerged and need to be urgently adressed such as data confidentiality and integrity. The amount of leaked information is also an important feature that needs to be considered and minimized in order to guarantee secrecy of wireless transmissions, \cite{8509094,8543573,BlochMatthieu1981-2011Ps:f}.

The concept of security started with Shannon's work, \cite{6769090}. These techniques are based on the assumption that the eavesdropper (Eve) has limited computational power capabilities. With the fast development in computing power devices, secret keys that were secure decades ago are nowadays more subject to successful brut-force attacks. Security is enhanced when the key length increases, resulting in more waste of resources. In addition, the key management processes become a real issue with the deployment of large-scale heterogeneous and decentralized networks involving different acces technologies, such as 5G networks. Finally, the emergence of power-limited, delay-sentive and processing-restricted wireless technologies, such as Internet Of Things (IoT), banking, health monitoring, vehicular communications, makes cryptography-based methods naturally unsuitable, \cite{8509094}.

To circumvent the aforementionned issues, physical layer security (PLS) has emerged as an effective way to enhance security of wireless communications, \cite{alves2012performance,yang2012physical,tran2015secrecy,8353879}. PLS classically takes benefit from unpredictable wireless channel characteristics (e.g., multipath fading, noise, dispersion, diversity) to improve security of communications against potential eavesdroppers without relying on computational complexity, i.e., the security is not affected if Eve has unlimited computing capabilities, \cite{9049811,snchez2020survey}. 

The starting point of PLS was exposed in 1975 by Wyner where he explained that a communication can be made secure, without sharing a secret key, when the wiretap channel of the eavesdropper is a degraded version, i.e., noisier, of the legitimate link (Bob), \cite{6772207}. This work was later extended to the broadcast channel in \cite{1055892}, and to the Gaussian channel in \cite{1055917}. 

It is of prime importance to evaluate the effectiveness of a PLS scheme by quantifying the degree of secrecy it can provide with a suitable metric. One of the most studied class of PLS metrics is the secrecy capacity. The information-theoretic secrecy-capacity is defined as the number of bits per channel use that can be reliably transmitted from a legitimate transmitter (Alice) to a legitimate receiver (Bob) while guaranteeing a negligible information leakage to the eavesdropper, \cite{bloch_barros_2011}.

 PLS can be achieved by increasing the signal-plus-interference to noise ratio (SINR) at Bob and decreasing the SINR at Eve. This can be done by designing a suitable channel-based adaptive transmission scheme, and/or by injecting an artificial noise (AN) signal to the data. These techniques can be implemented in the space, time and/or frequency domains, \cite{8509094,7762075,MELKI20191}.

Channel-based adaptation secrecy schemes were first introduced in \cite{4529264,4529282,4626059}. In these works, it was proven that positive secrecy rate (SR) can be obtained even if, on average, the channel between Alice and Bob is a degraded version of the one between Alice and Eve, by optimizing or adapting at the transmitter side the communication parameters. In doing so, the precoded signal is optimal for Bob's channel but not for Eve's one since they experience different fading. The concept of AN addition was first established in \cite{1558439,1605889,4543070}. The idea is to make Eve's channel condition artificially degraded by intentionnaly adding an AN signal to the transmitter data. This AN signal is designed in such a way not to degrade Bob's channel, therefore leading to a PLS enhancement, \cite{8509094}.

While many works implement these schemes with multiple antennas at the transmitter, using for instance frequency diverse array beamforming \cite{li_qiang_2018_1159254,8078202}, directional modulation (DM) \cite{5159486},  antenna subset modulation (ASM) \cite{6544472}, near-field direct antenna modulation (NFDAM) \cite{4684619,4523120}, spatial diversity \cite{5580113,7070667,8786136,5738303}, or waveform design \cite{9062307}, only few works perfom PLS using single-input single-output (SISO) systems \cite{li2013waveform,9003692,8093595,8093591,xu2018security,li2018artificial,li2017artificial,9049811,7475864,7041552}. SISO systems are indeed more suitable to resource-limited devices such as in IoT-type applications.   In \cite{li2013waveform}, a symbol waveform optimization technique in time-domain (TD) is proposed to reach a desired SINR at Bob with AN injection, under power constraint, when eavesdropper’s CSI is not known. Another approach to increase the SINR in SISO systems is time reversal (TR) pre-filtering. This has the advantage to be implemented with a simple precoder at the transmitter. TR achieves a focusing gain at the intended receiver position only, thereby naturally offering intrinsic anti-eavesdropping capabilities, \cite{9003692,oestges2005characterization}. TR is achieved by up/downsampling the signal in the TD. While the impact of the back-off rate (BOR), defined as the up/downsampling rate \cite{dubois2010use}, was studied in \cite{9003692,9049811}, limited non-optimal decoding capabilities were attributed to Eve. Another approach to provide security at the physical layer is the use of orthogonal frequency-division multiplexing (OFDM) scheme which can be implemented in time or frequency domain (FD). In \cite{8093595,8093591} FD OFDM schemes are presented consisting of subcarriers index selection. Only several subcarriers are used for data transmission depending on their channel gains.  In \cite{9062307}, the information-theoretic secrecy capacity of an Offset-QAM-based filterbank multicarrier (FBMC-OQAM) communication over a wiretap frequency selective channel is studied. The authors compare the secrecy capacity of the FMBC-OQAM modulation with a cyclic prefix-orthogonal frequency-division multiplexing (CP-OFDM) modulation.

To further enhance the secrecy, few works combine TD or FD precoding with AN injection, \cite{9049811,xu2018security,li2018artificial,li2017artificial,7475864,7041552}. In \cite{xu2018security,li2018artificial,li2017artificial}, TD TR precoders are presented. In these works, the AN is added either on all the channel taps or on a set of selected taps. While the condition for AN generation is given, its derivation is however not detailed. In \cite{9049811,7475864,7041552}, FD precoders using OFDM and AN injection are presented. In \cite{9049811}, the AN is injected in the null space of Bob but only limited decoding capacilities were attributed to Eve. In \cite{7475864,7041552}, the idea is to use several OFDM subcarriers for dummy data transmission, i.e., several subcarriers are used for data obfuscation. However, the encryption information must be shared between the transmitter and the legitimate receiver, leading to more processing needed at the receiver. In addition, the security is enhanced when more subcarriers are used for data obfuscation, at the expense of the data rate. Furthermore, it is assumed that Eve has no knowledge about the legitimate link.

In this paper, an original and novel FD TR precoder in SISO OFDM systems with AN addition is introduced to secure wireless communications. Indeed, TR can be equivalently implemented in FD by replicating and shifting the signal spectrum, \cite{8883213}. FD implementation has the advantage to be easily performed using OFDM. First results of this scheme were presented in \cite{9049811} where limited decoding capabilities were considered at Eve. In the following, three scenarios are investigated corresponding to the amount of channel's information Eve can obtain, which in turn depends on the handshake procedure.  In all the scenarios, Bob's CSI is fully known at Alice, using channel reciprocity in time division duplex (TDD) systems. An AN signal is designed in the FD in the presence of a passive eavesdropper whose instantaneous CSI is supposed unknown. The ergodic SR performance is derived with analytic models for the three investigated scenarios and compared to simulation results. It allows to determine the optimal amount of AN energy to inject in order to maximize the ergodic SR. Furthermore, it is proven that the investigated scheme can guarantee a desired SR for an infinite signal-to-noise ratio (SNR) at Eve. In addition, a power allocation technique, keeping into account the AN injection, is also derived in order to further enhance the SR. The proposed scheme uses only frequency diversity inherently present in multipath environments to achieve security. It can therefore be used in SISO systems and is then well-suited for resource-limited nodes such as encountered in IoT or vehicular communications for instance, \cite{9049811}. Finally, the OFDM implementation makes this approach compatible with LTE and 5G networks.

The reminder of this article is organized as follows: the communication and hanshake protocols are respectively exposed in Sections \ref{sec:communication-protocol} and \ref{sec:establishment}. Section \ref{sec:perf} presents a closed-form approximation of the amount of AN energy to be injected in order to maximize the SR, for the different decoding structures at Eve. The required SNR at Bob to guarantee a desired SR is derived, as a function of the communication parameters. Then, a waterfilling optimization procedure is assessed in order to further increase the communication SR. Theoretical and numerical results are shown in Section \ref{sec:simulation-results}. Section \ref{sec:conclusions} concludes the paper.

\textit{\textbf{Notation:}} the italic lower-case letter denotes a complex number. Greek letter corresponds to a scalar, the bold lower-case letter denotes a column vector. Bold upper-case letter corresponds to a matrix; $\textbf{I}_N$ is the $N \times N$ identity matrix; $(.)^{-1}$, $(.)^{*}$, $(.)^{H}$ are respectively the inverse, the complex conjugate and the Hermitian transpose operators; $\EX{.}$ is the expectation operator; $\module{.}$ is the modulus operator (element-wize modulus if matrix); $\odot$ is the element-wize (hadamard) product between two vectors of same dimension; $\vect{0}$ and $\vect{1}$ are respectively all-zero and all-one column vector of the right dimension.

\section{System Model}\label{sec:system-model}

\subsection{Communication Protocol}\label{sec:communication-protocol}
\begin{figure}[h!t]
	\centering
	\includegraphics[width=1\linewidth]{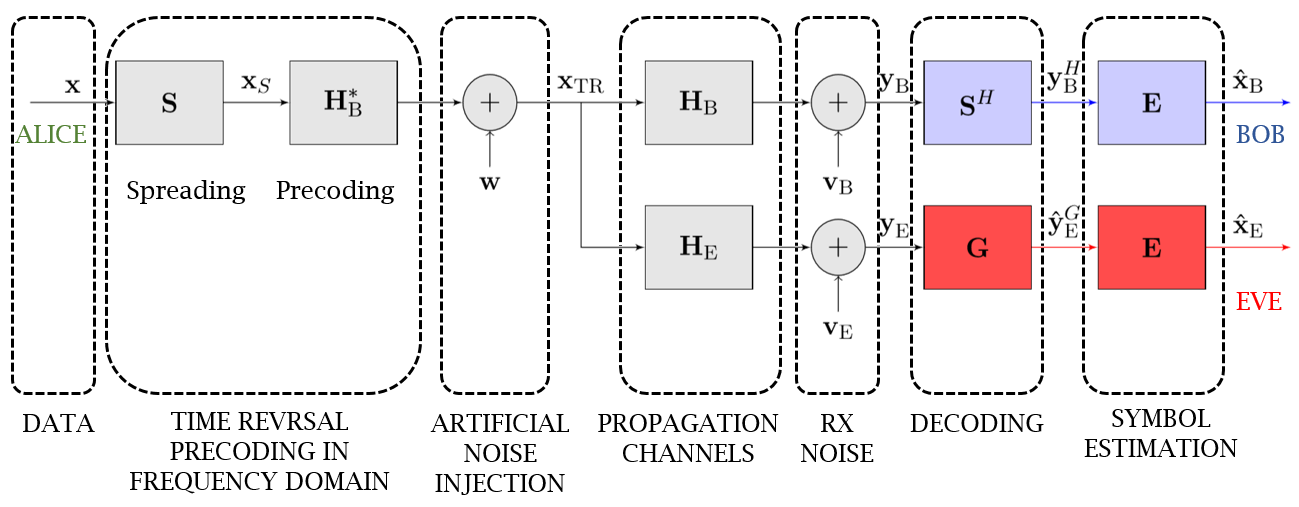}
	\caption{Communication scheme}
	\label{fig_com_scheme}
\end{figure}

In order to transmit secure data between Alice and Bob, the useful data is precoded and an AN signal $\w$ is added before transmission, as depicted in Fig.\ref{fig_com_scheme}.  
\\
The scheme consists in data transmission onto OFDM blocks with $Q$ subcarriers. Without loss of generality, it is considered that only one data block $\textbf{x}$ is sent and is composed of $N$ symbols $x_n$ (for $n = 0,..., N-1$, with $N\leq Q$). The symbol $x_n$ is  a zero-mean random variable (RV) with variance $\EX{|x_n|^2} = \sigma_x^2 = 1$, i.e., a normalized constellation is considered. The block is then spread in the FD by a factor $U = Q/N$, called back-off rate (BOR), thanks to the spreading matrix $\spread$ of size $Q\times N$. \textbf{S} is designed in such a way not to increase the PAPR, as suggested in \cite{4394231}.
\begin{equation}
\mat{S} = \frac{1}{\sqrt{U}} 
\begin{pmatrix}
\pm 1 & 0 & \hdots & 0 \\
0 & \pm 1 & \hdots & 0 \\
\vdots & & \ddots & \vdots \\
0 & 0 & \hdots & \pm 1 \\
& \vdots& \vdots& \\
\pm 1 & 0 & \hdots & 0 \\
0 & \pm 1 & \hdots & 0 \\
\vdots & & \ddots & \vdots \\
0 & 0 & \hdots & \pm 1
\end{pmatrix}
\label{eq:spread_mat}
\end{equation}
In doing so, each data symbol is transmitted onto $U$ different subcarriers with a spacing of $N$ subcarriers, introducing frequency diversity. The spread sequence is then precoded with the complex conjugate of Bob's channel $\HB^*$, before addition of the AN signal $\textbf{w}$  and transmission. 

The AN should not have any impact at Bob's position but should corrupt the data everywhere else since Alice does not have any information about Eve's instantaneous CSI, i.e., Eve is a passive node. Furthermore, this signal should not be guessed at the unintended positions to ensure the secure communication. With these considerations, the transmitted sequence becomes:
\begin{equation}
	\textbf{x}_{\text{TR}} = \sqrt{\alpha} \;\HB^*  \spread\; \textbf{x} +  \sqrt{1-\alpha} \; \w
	\label{eq:sym_rad_AN}
\end{equation} 
where $\alpha \in [0,1]$ defines the ratio of the total power sent dedicated to the useful signal, while ensuring that the total transmitted power remains constant, and equals to $1$ per transmitted symbol for any value of $\alpha$.

The channels between Alice and Bob ($\HB$) and between Alice and Eve ($\HE$) are $Q\times Q$ diagonal matrices whose elements are $h_{\text{B},q}$ and $h_{\text{E},q}$ (for $q = 0,...,Q-1$) and follow a zero-mean unit-variance complex normal distribution, i.e., their modulus follow a Rayleigh distribution. The overall channel energies are normalized to unity for each channel realization, i.e., $\mathbb{E}\left[\left|h_{\text{B},q}\right|^2\right]  = \mathbb{E}\left[\left|h_{\text{E},q}\right|^2\right] = 1, \; q = 0,...,Q-1$. The precoding matrix $\HB^*$ is also a diagonal matrix with elements $h_{\text{B},q}^*$. At Bob, a despreading operation is performed by applying $\spread^H$. It is assumed that Bob and Eve know the spreading sequence. Bob then applies a zero forcing (ZF) equalization, while Eve uses the best decoding structure \mat{G} she can, depending on the scenario, before applying a ZF equalization too, as explained in section \ref{sec:establishment}. A perfect synchronization is finally assumed at Bob and Eve positions.\\

\subsubsection{Artificial noise Design}\label{sec:artificial-noise-design}
In order not to have any impact at the intended position, the AN signal must satisfy the following condition:
\begin{equation}
	\textbf{A} \w \; = \; \textbf{0}
	\label{eq:an_cond}
\end{equation}
where $\textbf{A} = \spread^H\HB\; \in \C^{N\times Q}$. Condition (\ref{eq:an_cond}) ensures that $\w$ lies in the right null space of $\textbf{A}$. A singular value decomposition (SVD) of $\textbf{A}$ is performed, leading to:
\begin{equation}
	\textbf{A} = \textbf{U} 
	\begin{pmatrix}
		\Sigma \; \textbf{0}_{Q-N\times Q}
	\end{pmatrix}
	\begin{pmatrix}
		\textbf{V}_1^H \\
		\textbf{V}_2^H
	\end{pmatrix}
	\label{eq:an_svd}
\end{equation}
where $\textbf{U} \in \C^{N \times N}$ contains left singular vectors, $\Sigma \in \C^{N \times N}$ is a diagonal matrix containing singular values, $\textbf{V}_1 \in \C^{Q \times N}$ contains right singular vectors associated to non-zero singular values, and $\textbf{V}_2 \in \C^{Q \times Q-N}$ contains right singular vectors that span the right null space of $\textbf{A}$. Therefore, the AN signal can be expressed as:
\begin{equation}
	\w = \frac{\textbf{V}_2}{\sqrt{U |\mat{V_2}|^2}} \tilde{\w}
	\label{eq:an_w}
\end{equation}
which ensures that (\ref{eq:an_cond}) is satisfied for any arbitrary vector $\tilde{\w} \in \C^{Q-N \times 1}$. Since $Q = NU$, as soon as $U\geq 2$, there is a set of infinite possibilities to generate $\tilde{\w}$ and therefore the AN signal. In the following, it is assumed that $\tilde{\w}$ is a zero-mean circularly symmetric white complex Gaussian noise with covariance matrix $\EX{\tilde{\w}(\tilde{\w})^H} = \textbf{I}_{Q-N }$. The AN signal is then generated thanks to (\ref{eq:an_w}) with a normalization factor ensuring a total energy per symbol of 1.

\subsubsection{Received sequence at the intended position}
After despreading, the received sequence at Bob is: 
\begin{equation}
	\textbf{y}_{\text{B}}^H = \sqrt{\alpha} \; \spread^H \module{\HB}^2 \spread \textbf{x} \;  +  \;  \spread^H \textbf{v}_\text{B} 
	\label{eq:rx_bob_AN}
\end{equation}
where $\textbf{v}_\text{B}$ is the FD complex Additive White Gaussian Noise (AWGN) with noise's variance $\EX{|v_{\text{B},n}|^2}  = \sigma_{\text{V,B}}^2$ and covariance matrix $\EX{(\spread^H  \textbf{v}_\text{B}) (\spread^H \textbf{v}_\text{B})^H} = \sigma_{\text{V,B}}^2 \textbf{I}_N$. In (\ref{eq:rx_bob_AN}), each transmitted data symbol is affected by a real gain $ \frac{\sqrt{\alpha}}{U}\sum_{i=0}^{U-1} \left| h_{\text{B}, n + iN}\right|^2$ at the position of the legitimate receiver. This frequency diversity gain consequently increases the received useful signal power at Bob in fading environments and increases with the BOR value. Considering a fixed bandwidth, the TR focusing effect is enhanced for higher BOR's at the expense of the data rate. It is also observed that no AN contribution is present in (\ref{eq:rx_bob_AN}) since (\ref{eq:an_cond}) is respected. A ZF equalization is performed at the receiver leading to:
\begin{equation}
	\begin{split}
		\hat{\textbf{x}}_{\text{B}} &= \left( \sqrt{\alpha} \spread^H \module{\HB}^2 \spread \right)^{-1}  \left(\sqrt{\alpha}  \spread^H\module{\HB}^2 \spread \textbf{x}   +    \spread^H \textbf{v}_\text{B}\right) \\
		&= \textbf{x} + \left( \sqrt{\alpha} \spread^H \module{\HB}^2 \spread \right)^{-1} \spread^H \textbf{v}_\text{B}
	\end{split}
	\
	\label{eq:rx_bob_AN_eq}
\end{equation}
From (\ref{eq:rx_bob_AN_eq}), a perfect data recovery is possible in high SNR scenarios.

\subsubsection{Received sequence at the unintended position}
The received sequence at the eavesdropper position is given by:
\begin{equation}
	\textbf{y}_{\text{E}}^G = \sqrt{\alpha}  \textbf{G} \HE \HB^* \spread\textbf{x} + \sqrt{1-\alpha} \textbf{G} \HE \w + \textbf{G}  \textbf{v}_\text{E}
	\label{eq:rx_eve_an}
\end{equation}
where $\textbf{G}$ is a $N \times Q$ decoding matrix performed by Eve and $\textbf{v}_\text{E}$ is a complex AWGN. The nature of the decoding matrix is determined by the considered scenarios, which are presented in the next Section \ref{sec:establishment}. The noise variance is $\EX{|v_{\text{E,n}}|^2} = \sigma_{\text{V,E}}^2$.  The gain of the data component in (\ref{eq:rx_eve_an}) depends on \mat{G} and does not necessarily provide a SNR enhancement due to a TR effect. Similarly, the AN component does not necessarily cancel out, depending on \mat{G}. After ZF equalization, the estimated symbols are:
\begin{equation}
	\begin{split}
		\hat{\textbf{x}}_{\text{E}} =& \left(\textbf{G} \HE \HB^* \spread \right)^{-1}
		\left( \sqrt{\alpha} \textbf{G} \HE \HB^* \spread \textbf{x} +   \sqrt{1-\alpha} \textbf{G} \HE \w  +  \textbf{G}  \textbf{v}_\text{E}  \right) \\
		=& \sqrt{\alpha}\textbf{x} + \sqrt{1-\alpha} \left(\textbf{G} \HE \HB^* \spread \right)^{-1}  \textbf{G} \HE \w + \left(\textbf{G} \HE \HB^* \spread \right)^{-1}  \textbf{G} \textbf{v}_\text{E}
	\end{split}
	\label{eq:rx_an_eve_eq}
\end{equation}
Equation (\ref{eq:rx_an_eve_eq}) shows that the addition of AN in the FD TR SISO OFDM communication can secure the data transmission. The degree of security depends on \mat{G} and the amount of data energy, $\alpha$, that is injected into the communication, with respect to the amount of AN energy injected (via $1-\alpha$), as explained in Section \ref{sec:perf}. It is to be noted that, since $\w$ is generated from an infinite set of possibilities, even if Eve knows its equivalent channel $\HE\HB^*$ and the spreading sequence, she cannot estimate the AN signal  to try retrieving the data.

\subsection{Handshake Protocol and related assumptions}\label{sec:establishment}
Prior to the secure data transmission between Alice and Bob, a handshake protocol must take place. Depending on it, Eve may obtain different degrees of information regarding the channels, which leads to different decoding capabilities and so, different security performance. PLS performance highly depends on the availability of CSI at the communication parties. It is assumed that Alice knows Bob CSI but does not know Eve CSI  who is assumed to be an external passive node of the network that tries to eavesdrop the data. Furthermore, Bob and Eve CSI's are considered spatially independent.

In this paper, a Fast Fading (FF) TDD communication is considered. In doing so, three different decoding schemes are investigated at Eve depending on weither Alice or Bob first initiates the secure communication. The FF hypothesis means that each OFDM block sent by Alice experiences a different channel realization. It results in an impossibility for Eve to learn some parameters from the communication, such as the AN variance, since Bob's channel varies too rapidly and has to be frequently re-estimate by Alice. 

The first two scenarios occur when Bob first requests to Alice for secure communication. In both cases, Bob transmits a pilot to Alice allowing her to estimate Bob's channel.

\begin{figure}[!htb]
	\centering
	\includegraphics[width=.9\linewidth]{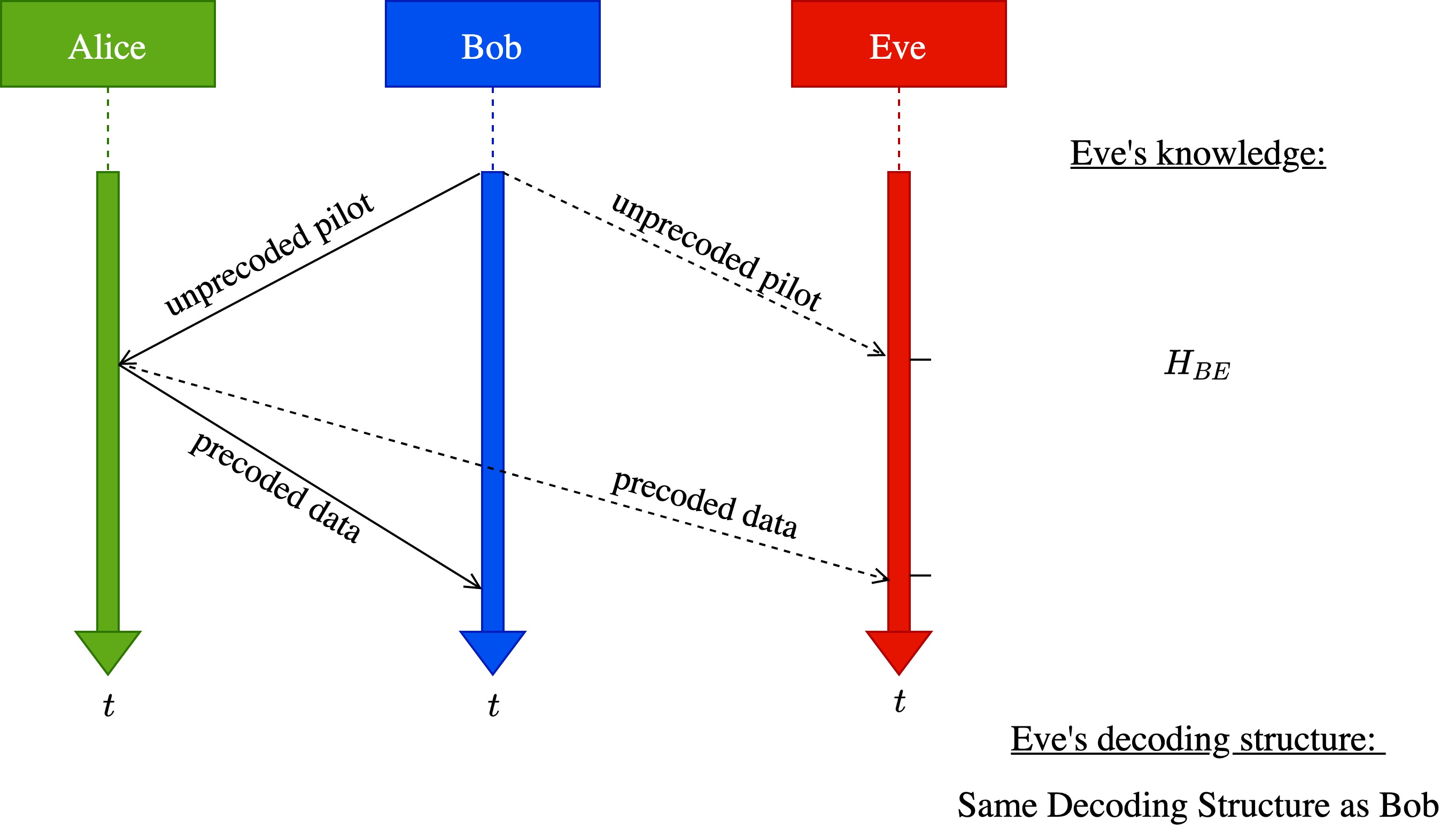}
	\caption{FF TDD, Bob initiates the communication, no pilot sent}
	\label{fig_ff_tdd_b_no_pilot}
\end{figure}

If Alice only transmits precoded data to Bob, Eve is not able to know anything but $\textbf{H}_{BE}$, the channel between Bob and Eve. In that situation, she cannot do better but to implement the same decoding structure as Bob, denoted by the abbreviation \textit{SDS}. In that scenario, Eve only despreads the received sequence. This situation is presented in Fig.\ref{fig_ff_tdd_b_no_pilot}.

\begin{figure}[!htb]
	\centering
	\includegraphics[width=.95\linewidth]{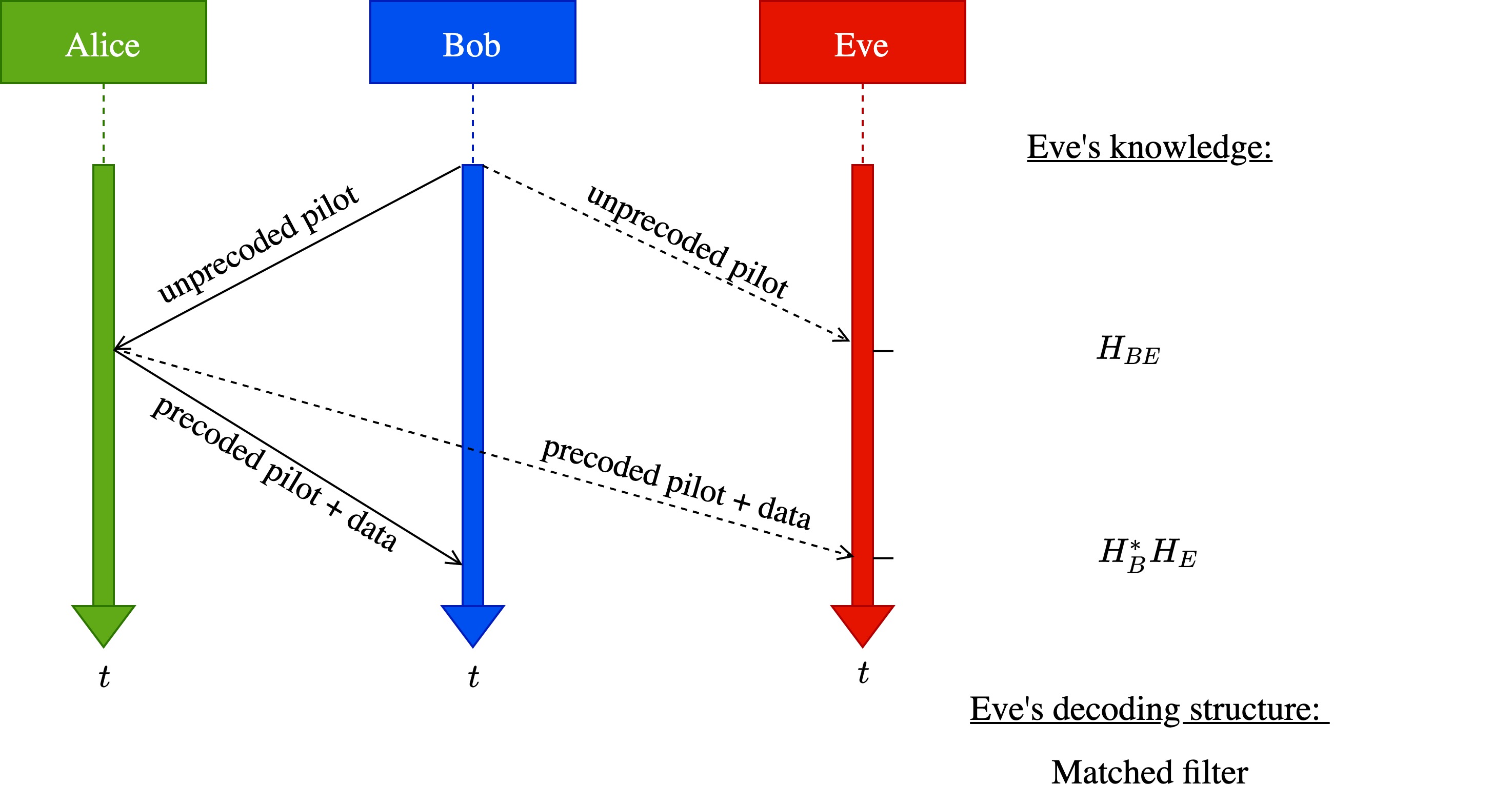}
	\caption{FF TDD, Bob initiates the communication, pilots sent}
	\label{fig_ff_tdd_b_pilot}
\end{figure}

However, if Alice sends a precoded pilot in addition to the  precoded data to Bob, Eve is then able to evaluate her equivalent channel $\HB^*\HE$, and therefore to implement a matched filtering decoding structure, denoted by \textit{MF}. This is depicted in Fig.\ref{fig_ff_tdd_b_pilot}

\begin{figure}[!htb]
	\centering
	\includegraphics[width=.85\linewidth]{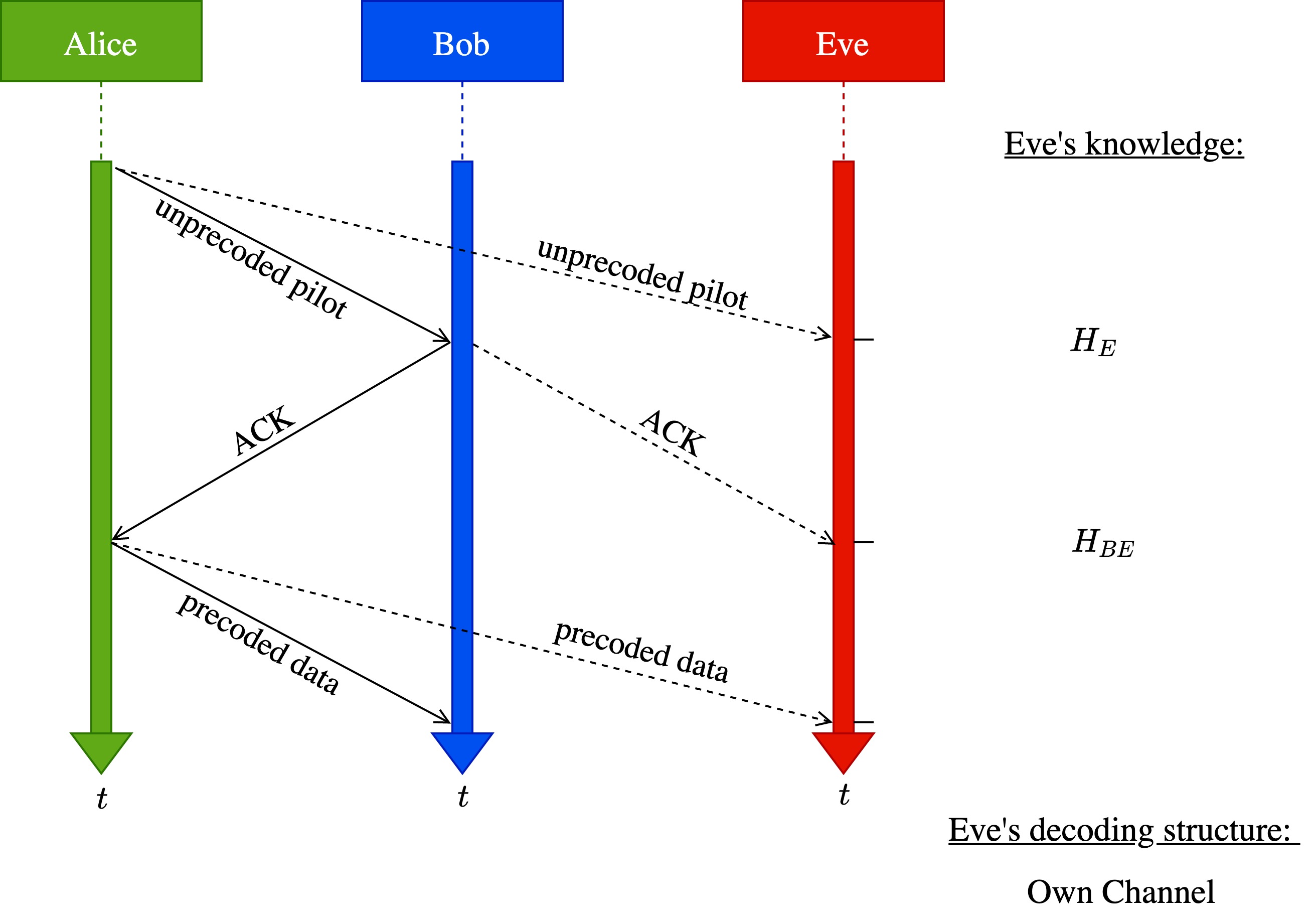}
	\caption{FF TDD, Alice initiates communication}
	\label{fig_ff_tdd_a}
\end{figure}

The third investigated situation arises when Alice asks first to Bob for secure commnication, as depicted in Fig.\ref{fig_ff_tdd_a}. In this configuration, she sends a pilot to Bob allowing Eve to estimate her own channel frequency response (CFR) $\HE$. From that, Bob acknowledges to Alice without needing to send his own channel estimate $\HB$, thanks to the channel reciprocity property in TDD systems. Here, Alice estimates $\HB$ thanks to an a prirori known ACK. Finally, Alice sends precoded data without pilot to Bob. From the FF assumption, Eve cannot learn the precoding performed by the transmitter. In this configuration, Eve implements  a decoding structure that takes benefit of her own channel knowledge, denoted by \textit{OC}. 

In the following section, the three decoding schemes are studied. Analytical models of their performance are derived and discussed.

%
%
%


\section{Performance Assessments}
\label{sec:perf}
The classical metric used to evaluate the degree of secrecy in a communication in the PLS field is the secrecy channel capacity (SC). The SC is defined as the maximum transmission rate that can be supported by the legitimate receiver's channel while ensuring the impossibility for the eavesdropper to retrieve the data, \cite{7348007}. In the ergodic sense, it can be expressed as:
\begin{equation}
\begin{split}
C_S &=  \EX{\log_2{\left(1+\gamma_B\right)} - \log_2{\left(1+\gamma_E\right)}}^+ 
\end{split}
\label{eq:SR}
\end{equation}
where $\left[x\right]^+ = \text{max}(x,0)$,  $\gamma_B$ and $\gamma_E$ being respectively the SINR at Bob and Eve's positions.  It was shown in \cite{8418798}, Lemma 1, that an achievable ergodic secrecy rate (SR), i.e., a positive rate smaller than or equal to the SC, is given by:
\begin{equation}
	\begin{split}
		R_S& = \left[\EX{\log_2(1+\gamma_B)-\log_2(1+\gamma_E)}\right]^+ \\
		& \approx \left[ \log_2(1 + \EX{\gamma_B}) - \log_2(1 + \EX{\gamma_E})\right]^+
	\end{split}
	\label{eq:SR2}
\end{equation}
As it is observed in simulations, the bound (\ref{eq:SR2}) is tight, leading to an accurate approximation. To estimate the SR of the communication, the analytical expressions of Bob and Eve ergodic SINR's are derived in the following sections.


\subsection{Hypothesis}
In order to obtain the analytical models, the following assumptions are considered:
\begin{itemize}
	\item The data and noise are independent of each other. 
	\item $h_{B,i} \independent h_{B,j}, \forall i \neq j$, i.e., no frequency correlation between Bob's channel subcarriers
	\item $h_{E,i} \independent h_{E,j}, \forall i \neq j$, i.e., no frequency correlation between Eve's channel subcarriers\footnote{Thanks to the design of the spreading matrix, the $U$ subcarriers composing one symbol are spaced by $N = Q/U$ subcarriers. If this distance is larger than the coherence bandwidth of the channel, the assumption holds. This usually occurs in rich multipath environments and for sufficiently large bandwidths and moderate BOR values.}.
	\item $h_{B,i} \independent h_{E,j}, \forall i,j$, i.e., Bob and Eve are sufficiently spaced leading to no spatial correlation between them.
\end{itemize}


\subsection{SINR determination}
In this section, the ergodic SINR for transmitted symbol $n, \; n = 0,...,N-1$ at Bob and Eve positions are derived, depending on the investigated scenario, i.e., on the handshake procedure.

\subsubsection{At the intended position}
At Bob, a simple despreading operation is performed. Thanks to the precoding at the transmitter side, every received data symbol is affected by a real gain, as expressed in (\ref{eq:rx_bob_AN}). The ergodic SINR for transmitted symbol $n$ is given by:
\begin{equation}
\begin{split}
\EX{\gamma_{B,n}} &= \EX{ \frac{  \left| \sqrt{\alpha} B_{1,n} x_n \right|^2  }{  \left| B_{2,n} \right|^2} }  = \alpha \EX{\left| B_{1,n}  x_n\right|^2}  \EX{\frac{1}{\left| B_{2,n} \right|^2}}  \\
& \geq  \frac{\alpha \EX{  \left| B_{1,n}  x_n\right|^2 } }{\EX{ \left| B_{2,n} \right|^2 }} =  \frac{\alpha \EX{ \left|B_{1,n}  \right|^2 } \EX{ \left| x_n \right|^2 } }{\EX{ \left| B_{2,n} \right|^2 }}
\label{eq:RV_sinr_b}
\end{split}
\end{equation}
where $B_{1,n} = \frac{\sqrt{\alpha}}{U}\sum_{i=0}^{U-1} \left| h_{\text{B}, n + iN}\right|^2$, $x_n$ is the $n^{\text{th}}$ data symbol at Bob, and $B_{2,n} = \frac{1}{\sqrt{U}}\sum_{i=0}^{U-1} \left| v_{\text{B}, n + iN}\right|$ is the $n^{\text{th}}$ noise symbol component at Bob and where it is observed that $B_{1,n} \independent x_n \independent B_{2,n}$.\\
As detailed in \ref{sec:data-term-app} and \ref{sec:awgn-term-app}, the components can respectively be derived as:
\begin{equation}
	\begin{split}
	\EX{|B_{1,n}|^2} =& \frac{\alpha (U+1)}{U}
	\end{split}
	\label{eq:appA:data_bob}
\end{equation}
\begin{equation}
	\begin{split}
	\EX{|B_{2,n}|^2} &= \sigma^2_{\text{V,B}}
	\end{split}
	\label{eq:appA:noise_bob}
\end{equation}

From (\ref{eq:RV_sinr_b}), (\ref{eq:appA:data_bob}) and (\ref{eq:appA:noise_bob}), the ergodic SINR for particular symbol $n$ at the intended position is thus given by:
\begin{equation}
\EX{\gamma_{B,n}} \geq \frac{\alpha \;(U+1)}{U \; \sigma_{\text{V,B}}^2}
\label{sinr_bob}
\end{equation}
As a reminder, the transmitted energy per symbol is equal to 1. It has been observed in simulations than the lower-bound (\ref{sinr_bob}) is tight enough to be used as an approximation for the averaged SINR at the intended position.

\subsubsection{At the unintended position}
At the unintended position, the received signal before ZF equalization is given by (\ref{eq:rx_eve_an}). Let's introduce $\textbf{E}_1^G = \sqrt{\alpha}  \textbf{G} \HE \HB^* \spread\textbf{x} $, $\textbf{E}_2^G = \textbf{G}  \textbf{v}_\text{E}$ and $\textbf{E}_3^G = \sqrt{1-\alpha} \textbf{G} \HE \w$ being respectively the data component, the noise component, and the AN component of the received signal at Eve  for a particular decoding structure $\textbf{G}$. Using the Jensen's inequality, an approximation of a lower-bound of the averaged SINR of the symbols $n$ at the unintended position can be derived as\footnote{Neglecting the covariance between $\left|E_{1,n}^G\right|^2$ and $\left| E_{2,n}^G + E_{3,n}^G \right|^2$, as  done in the first line of (\ref{eq:expected_sinr_eve}), makes the nature of the bound, i.e., lower or upper, obtained for $\EX{\gamma_{E,n}^G}$ uncertain. However, it has been observed by simulations that it remains very tight and a lower one for all considered scenarios.}:

\begin{equation}
\begin{split}
\EX{\gamma_{E,n}^G} &= \EX{  \frac{ \left| E_{1,n}^G \right|^2  }{ \left| E_{2,n}^G + E_{3,n}^G \right|^2 } }  \approx  \EX{ \left| E_{1,n}^G \right|^2 }  \EX{ \frac{1}{ \left| E_{2,n}^G + E_{3,n}^G \right|^2} }  \\
& \gtrapprox \frac{\EX{   \left| E_{1,n}^G \right|^2  } }{\EX{ \left| E_{2,n}^G + E_{3,n}^G \right|^2  }} =  \frac{\EX{  \left| E_{1,n}^G \right|^2  } }{\EX{  \left| E_{2,n}^G \right|^2  } +  \EX{  \left|E_{3,n}^G \right|^2  }}
\label{eq:expected_sinr_eve}
\end{split}
\end{equation}
where $E_{1,n}^G$, $E_{2,n}^G$ and $E_{3,n}^G$ are respectively the data, noise and AN $n^{\text{th}}$ symbol components of the received signal at Eve's position, for a particular decoding structure \mat{G}. The expression of the SINR at Eve depends on the receiving structure $\textbf{G}$ whose design depends on the amount of knowledge Eve can obtain. The expression (\ref{eq:expected_sinr_eve}) is therefore derived for  the three considered scenarios.

\paragraph{SDS Decoder}
\label{sec:same-decoding-strucure-as-bob}
This scenario corresponds to the situation presented in Fig.\ref{fig_ff_tdd_b_no_pilot} where Eve can only obtain the knowledge of $\mat{H}_{\text{BE}}$, which is of no help. The optimal decoding structure at Eve is therefore $\textbf{G}=\spread^H$. In that case, the received sequence becomes:
\begin{equation}
\textbf{y}_{\text{E}}^{SDS} = \sqrt{\alpha} \spread^H \HE \textbf{H}^*_{\text{B}} \spread\textbf{x} + \sqrt{1-\alpha} \spread^H \HE \w  +  \spread^H  \ve 
\label{eq:rx_eve_filt0}
\end{equation}
The symbol components can be written as:
\begin{equation}
\begin{split}
E_{1,n}^{SDS} &= \sqrt{\alpha}\frac{1}{U}\sum_{i=0}^{U-1}  h_{\text{E}, n + iN} h^*_{\text{B}, n + iN} \\
E_{2,n}^{SDS} &= \frac{1}{\sqrt{U}}\sum_{i=0}^{U-1}  v_{\text{E}, n + iN}\\
E_{3,n}^{SDS} &= \sqrt{1-\alpha}\frac{1}{\sqrt{U}}\sum_{i=0}^{U-1}  h_{\text{E}, n + iN}w_{n + iN}
\end{split}
\end{equation}
As detailed in  \ref{sec:data-term-app-1},  \ref{sec:awgn-term-app-1}, and \ref{sec:an-term-app-1}, the components can respectively be expressed as:
\begin{equation}
	\begin{split}
	\EX{|E_{1,n}^{SDS}|^2}&= \frac{\alpha}{U}
	\end{split}
	\label{eq:appA:data_eve_filt0}
\end{equation} 
\begin{equation}
	\begin{split}
	\EX{|E_{2,n}^{SDS}|^2} &=\sigma^2_{\text{V,E}}
	\end{split}
	\label{eq:appA:noise_eve_filt0}
\end{equation} 
\begin{equation}
	\begin{split}
	\EX{|E_{3,n}^{SDS}|^2}  &= \frac{1-\alpha}{U}
	\end{split}
	\label{eq:appA:an_eve_filt0}
\end{equation}
From (\ref{eq:expected_sinr_eve}), (\ref{eq:appA:data_eve_filt0}), (\ref{eq:appA:noise_eve_filt0}), and (\ref{eq:appA:an_eve_filt0}), the ergodic SINR for particular symbol $n$ when Eve has the same capabilities as Bob is given by:
\begin{equation}
\EX{\gamma_{E,n}^{SDS}} \gtrapprox \frac{\frac{\alpha}{U}}{\sigma^2_{\text{V,E}}+\frac{1-\alpha}{U}}
\label{eq:sinr_eve_filt0}
\end{equation}
Relatively low performances at Eve are expected with this decoding structure since the despreading operation does not coherently sum up the received symbol components. No frequency diversity gain is consequently achieved, leading to suboptimal decoding performances.

\paragraph{MF Decoder}
In this scenario, depicted in Fig.\ref{fig_ff_tdd_b_pilot}, Eve obtains the knowledge of $\HB^*\HE$, which allows her to implement a matched filtering decoding structure $\textbf{G} = \spread^H \HB\HE^*$. The received signal is therefore given by:
\begin{equation}
\begin{split}
\textbf{y}_{\text{E}}^{MF} =& \sqrt{\alpha} \spread^H \module{\HE}^2 \module{\HB}^2 \spread\textbf{x} +  \sqrt{1-\alpha} \spread^H \HB\module{\HE}^2 \w\\
&+  \spread^H  \textbf{H}^*_E \textbf{H}_B \ve
\end{split}
\label{eq:rx_eve_filt1}
\end{equation}
In this scenario, the symbol components become:
\begin{equation}
\begin{split}
E_{1,n}^{MF} &= \frac{\sqrt{\alpha}}{U}\sum_{i=0}^{U-1}  \left|h_{\text{E}, n + iN}\right|^2 \left|h_{\text{B}, n + iN}\right|^2 \\
E_{2,n}^{MF} &= \frac{1}{\sqrt{U}}\sum_{i=0}^{U-1} h^*_{\text{E}, n + iN} h_{\text{B}, n + iN} v_{\text{E}, n + iN}\\
E_{3,n}^{MF} &=\frac{ \sqrt{1-\alpha}  }{\sqrt{U}}\sum_{i=0}^{U-1}    h_{\text{\text{B}}, n + iN} \left|h_{\text{E}, n + iN}\right|^2w_{n + iN}
\end{split}
\end{equation}
As detailed in  \ref{sec:data-term-app-2},  \ref{sec:awgn-term-app-2}, and \ref{sec:an-term-app-2}, the components can respectively be derived as:
\begin{equation}
\EX{|E_{1,n}^{MF}|^2} =  \frac{\alpha (U+3)}{U}
\label{eq:data_eve_filt1}
\end{equation}
\begin{equation}
	\begin{split}
	\EX{|E_{2,n}^{MF}|^2} &= \sigma^2_{\text{V,E}}
	\end{split}
	\label{eq:noise_eve_filt1}
\end{equation}
\begin{equation}
	\EX{|E_{3,n}^{MF}|^2} = \frac{1-\alpha}{U+1}
	\label{eq:an_eve_filt1}
\end{equation}
From (\ref{eq:expected_sinr_eve}), (\ref{eq:data_eve_filt1}), (\ref{eq:noise_eve_filt1}), and (\ref{eq:an_eve_filt1}),  the ergodic SINR for particular symbol $n$ when Eve matched filters the received sequence is given by:
\begin{equation}
\EX{\gamma_{E,n}} \gtrapprox \frac{\alpha \frac{U+3}{U}}{\sigma^2_{\text{V,E}} + \frac{1-\alpha}{U+1}}
\label{eq:sinr_eve_filt1}
\end{equation}
The numerator in (\ref{eq:sinr_eve_filt1}) is about U times larger than in (\ref{eq:sinr_eve_filt0}) thanks to a frequency diversity gain.

\paragraph{OC Decoder}
\label{sec:own-channel-knowledge}
This situation is shown in Fig.\ref{fig_ff_tdd_a} where Eve can decode the data thanks to $\textbf{G} = \spread^H \HE^*$. The received sequence is:
\begin{equation}
	\textbf{y}_{\text{E}}^{OC} = \sqrt{\alpha} \spread^H \module{\HE}^2 \HB^* \spread\textbf{x} +  \sqrt{1-\alpha} \spread^H \module{\HE}^2 \w  +  \spread^H  \HE^*  \ve
	\label{eq:rx_eve_filt5}
\end{equation}
With this decoding structure, the received symbol components are defined as:
\begin{equation}
\begin{split}
E_{1,n}^{OC} &= \frac{\sqrt{\alpha}}{U}\sum_{i=0}^{U-1}  \left|h_{\text{E}, n + iN}\right|^2  h_{\text{B}, n + iN}^* \\
E_{2,n}^{OC} &= \frac{1}{\sqrt{U}}\sum_{i=0}^{U-1} h^*_{\text{E}, n + iN}  v_{\text{E}, n + iN}\\
E_{3,n}^{OC} &=\frac{ \sqrt{1-\alpha}  }{\sqrt{U}}\sum_{i=0}^{U-1}   \left|h_{\text{E}, n + iN}\right|^2 w_{n + iN}
\end{split}
\end{equation}
As detailed in  \ref{sec:data-term-app-3},  \ref{sec:awgn-term-app-3}, and \ref{sec:an-term-app-3}, the components can respectively be expressed as:
\begin{equation}
\EX{|E_{1,n}^{OC}|^2} = \frac{2\alpha}{U}
\label{eq:data_eve_filt5}
\end{equation}
\begin{equation}
\begin{split}
\EX{|E_{2,n}^{OC}|^2} &= \sigma^2_{\text{V,E}}
\end{split}
\label{eq:noise_eve_filt5}
\end{equation}
\begin{equation}
\begin{split}
\EX{|E_{3,n}^{OC}|^2}  &=  \frac{2(1-\alpha)}{U}
\end{split}
\label{eq:an_eve_filt5}
\end{equation}
From (\ref{eq:expected_sinr_eve}), (\ref{eq:data_eve_filt5}), (\ref{eq:noise_eve_filt5}), and (\ref{eq:an_eve_filt5}),  the ergodic SINR for particular symbol $n$ when Eve knows her own channel is given by:
\begin{equation}
\EX{\gamma_{E,n}^{OC}} \gtrapprox \frac{\frac{\alpha }{U}}{\frac{\sigma^2_{\text{V,E}}}{2} + \frac{1-\alpha}{U}}
\label{eq:sinr_eve_filt5}
\end{equation}
One can observe that (\ref{eq:sinr_eve_filt5}) is very similar to (\ref{eq:sinr_eve_filt0}). In particular, (\ref{eq:sinr_eve_filt5}) leads to slightly higher SINR values at Eve than (\ref{eq:sinr_eve_filt0}), especially at high $\sigma^2_{\text{V,E}}$ and high $\alpha$.

\subsection{Optimal amount of data energy to inject}
\label{subsec:best_alpha}
It has to be pointed out that lower bounds of the SINR at Bob and Eve were determined for the three investigated scenarios. From simulations, the closed form approximated SINR lower bounds, derived in (\ref{sinr_bob}),  (\ref{eq:sinr_eve_filt0}), (\ref{eq:sinr_eve_filt1}), and (\ref{eq:sinr_eve_filt5}), are observed to be very tight and are therefore used in the remaining as an approximation. By doing so, an analytical expression of the SR can be determined using (\ref{eq:SR2}) as a function of $\alpha$. It is therefore straightforward to determine the amount of data energy to inject in the communication, with respect to AN, in order to maximize the ergodic SR.
\subsubsection{SDS Decoder}
 With (\ref{eq:SR}), (\ref{sinr_bob}), and (\ref{eq:sinr_eve_filt0}), the SR becomes:
\begin{equation}
R_s^{SDS} \approx \log_2 \left( 1 +  \frac{\alpha \;(U+1)}{U \; \sigma_{\text{V,B}}^2} \right) - \log_2\left( 1 + \frac{\frac{\alpha}{U}}{\sigma^2_{\text{V,E}}+\frac{1-\alpha}{U}}\right)
\label{eq:SR_anal2_decod_0}
\end{equation}
It can be shown that the SR is maximized for:
\begin{equation}
\alpha_{\text{opt}}^{SDS} = \frac{(U+1)(U\sigma_{\text{V,E}}^2 + 1)- U\sigma_{\text{V,B}}^2}{2(U+1)}
\label{eq:optimal_alpha_decod_0}
\end{equation}

\subsubsection{MF Decoder}
With (\ref{eq:SR}), (\ref{sinr_bob}), and (\ref{eq:sinr_eve_filt1}), the SR is expressed as:
\begin{equation}
R_s^{MF} \approx \log_2 \left( 1 +  \frac{\alpha \;(U+1)}{U \; \sigma_{\text{V,B}}^2} \right) - \log_2\left( 1 +  \frac{\alpha \frac{U+3}{U}}{\sigma^2_{\text{V,E}} + \frac{1-\alpha}{U+1}}\right)
\label{eq:SR_anal2_decod_1}
\end{equation}
By introducing $T_1 = U+1$, $T_2 = (U+1)^2\sigma_{\text{V,E}}^2 + (U+1) - U\sigma_{\text{V,B}}^2$, $T_3 = U(U+1)\sigma_{\text{V,B}}^2\sigma_{\text{E}}^2 + U \sigma_{\text{V,B}}^2$, and $T_4=(U+1)(U+3)\sigma_{\text{V,B}}^2-U\sigma_{\text{V,B}}^2$, the optimal amount of data energy to transmit is: 
\begin{equation}
\alpha_{\text{opt}}^{MF} = \frac{\pm\sqrt{T_1^2 T_3^2 \; + \; T_1 T_2 T_3 T_4 \; - \; T_1 T_3 T_4^2} \; - \; T_1 T_3}{T_1 T_4}
\label{eq:optimal_alpha_decod_1}
\end{equation}
where only the positive root is solution since $\alpha \in [0,1]$.

\subsubsection{OC Decoder}
With (\ref{eq:SR}), (\ref{sinr_bob}) and (\ref{eq:sinr_eve_filt5}), the SR expression is given by:
\begin{equation}
R_s^{OC} \approx \log_2 \left( 1 +  \frac{\alpha \;(U+1)}{U \; \sigma_{\text{V,B}}^2} \right) - \log_2\left( 1 +  \frac{\frac{\alpha }{U}}{\frac{\sigma^2_{\text{V,E}}}{2} + \frac{1-\alpha}{U}}\right)
\label{eq:SR_anal2_decod_5}
\end{equation}
Therefore, the optimal amount of data energy to inject is given by:
\begin{equation}
\alpha_{\text{opt}}^{OC} = \frac{(U+1)(2+U\sigma_{\text{V,E}}^2) - 2U\sigma_{\text{V,B}}^2 }{4(U+1)}
\label{eq:optimal_alpha_decod_5}
\end{equation}

\subsection{Required SNR at Bob to guarantee a desired SR}
 \label{sec:required-snr-at-bob-for-a-targetted-sr}
With the closed-form approximations of the SR (\ref{eq:SR_anal2_decod_0}), (\ref{eq:SR_anal2_decod_1}), and (\ref{eq:SR_anal2_decod_5}), it is possible to determine the SNR at Bob and the amount of data energy $\alpha$ that guarantees a given SR, as a function of the communication parameters. Let's introduce $\Delta$ being the targetted SR in bit per channel use, $\delta_B^{SDS}, \delta_B^{MF}$, and $\delta_B^{OC}$ being respectively Bob's required SNR for the first, second and third investigated scenario. Remembering that $\sigma^2_{\text{V,B}} = \frac{1}{U\delta_{B}}$, the SNR can be found as:
\begin{equation}
\delta_{B}^{SDS} = \frac{2^\Delta (U\sigma_{\text{V,E}}^2+1) - (U\sigma_{\text{V,E}}^2+1-\alpha) }{\alpha(U+1)(U\sigma_{\text{V,E}}^2+1-\alpha)}
\label{eq:snr_calibration_bob_SDS}
\end{equation}
\begin{equation}
\resizebox{.435 \textwidth}{!} 
{
	$\delta_{B}^{MF} = \frac{2^\Delta \left[U(U+1)\sigma_{\text{V,E}}^2 + \alpha(U+1)(U+3)+U(1-\alpha) \right] - \left[ U(U+1)\sigma_{\text{V,E}}^2+U(1-\alpha)\right] }{\alpha U(U+1)\left[(U+1)\sigma_{\text{V,E}}^2+(1-\alpha)\right]}$
}
	\label{eq:snr_calibration_bob_MF}
\end{equation}
\begin{equation}
\delta_{B}^{OC} = \frac{2^\Delta  (U\sigma_{\text{V,E}}^2+2) - (U\sigma_{\text{V,E}}^2+2-2\alpha) }{ \alpha(U+1)(U\sigma_{\text{V,E}}^2+2-2\alpha)}
\label{eq:snr_calibration_bob_OC}
\end{equation}
Equations (\ref{eq:snr_calibration_bob_SDS}), (\ref{eq:snr_calibration_bob_MF}), and (\ref{eq:snr_calibration_bob_OC})  give the required SNR at Bob to target SR $=\Delta$, as a function of the BOR $U$ and the noise level at Eve $\sigma_{\text{V,E}}^2$. \\
To be able to guarantee SR $=\Delta$, one has to consider Eve's SNR $= \infty$ in equations (\ref{eq:snr_calibration_bob_SDS}), (\ref{eq:snr_calibration_bob_MF}), and (\ref{eq:snr_calibration_bob_OC}) as Eve's SNR is not known to Alice.\\
Let's introduce $\delta_E$ as Eve's SNR. One finds $\sigma^2_{\text{V,E}} = \frac{1}{U\delta_{E}} = 0$. Introducing $ \sigma_{\text{V,E}}^2 = 0$ in (\ref{eq:snr_calibration_bob_SDS}), (\ref{eq:snr_calibration_bob_MF}), and (\ref{eq:snr_calibration_bob_OC}), and denoting $\delta_{B,\infty}^{SDS}, \delta_{B,\infty}^{MF}, \; \text{and}\;  \delta_{B,\infty}^{OC}$ respectively as Bob's required SNR to guarantee SR $=\Delta$ for the first, second, and third investigated scenario when $\delta_E = \infty$, it comes:
\begin{align}
\delta_{B,\infty}^{SDS} &= \frac{\alpha + 2^\Delta - 1}{(-\alpha^2 + \alpha)(U+1)}
\label{eq:snr_calibration_bob_SDS_infinite} \\
\delta_{B,\infty}^{MF} &= \frac{\alpha\left[  2^\Delta(U+1)(U+3) - U(2^\Delta-1)  \right] + U(2^\Delta-1)}{(-\alpha^2 + \alpha)U(U+1)} 
\label{eq:snr_calibration_bob_MF_infinite}\\
\delta_{B,\infty}^{OC} &= \frac{\alpha + 2^\Delta - 1}{(-\alpha^2 + \alpha)(U+1)} = \delta_{B,\infty}^{SDS}
\label{eq:snr_calibration_bob_OC_infinite}
\end{align}
Equations  (\ref{eq:snr_calibration_bob_SDS_infinite}), (\ref{eq:snr_calibration_bob_MF_infinite}), and (\ref{eq:snr_calibration_bob_OC_infinite}) are convex expressions that can be minimized as a function of $\alpha$. Let's denote $\alpha_{\infty}^{SDS}, \alpha_{\infty}^{MF}, \; \text{and}\; \alpha_{\infty}^{OC}$ as the amount of data energy to inject, with respect to AN, in order to guarantee a desired communication SR when $\delta_E = \infty$, respectively for the first, second and third scenario. It can be shown that:
\begin{align}
	\alpha_{\infty}^{SDS} &= \sqrt{(2^\Delta -1)^2 + (2^\Delta -1)} - (2^\Delta -1)
	\label{eq:alpha_infty_SDS} \\
	\alpha_{\infty}^{MF} &= \frac{-2A_2+\sqrt{4A_1A_2 + 4A_2^2}}{2A_1} 
	\label{eq:alpha_infty_MF} \\
	\alpha_{\infty}^{OC} &= \sqrt{(2^\Delta -1)^2 + (2^\Delta -1)} - (2^\Delta -1) = \alpha_{\infty}^{SDS}
	\label{eq:alpha_infty_OC}
\end{align}
where $A_1 = 2^\Delta(U+1)(U+3)-U(2^\Delta-1)$, $A_2 = U(2^\Delta-1)$.\\
By replacing the values of $\alpha$ in (\ref{eq:snr_calibration_bob_SDS_infinite}), (\ref{eq:snr_calibration_bob_MF_infinite}), and (\ref{eq:snr_calibration_bob_OC_infinite}) respectively by (\ref{eq:alpha_infty_SDS}), (\ref{eq:alpha_infty_MF}), and (\ref{eq:alpha_infty_OC}), the expressions of Bob's SNR to ensure SR $=\Delta$ are found.

\subsection{Secrecy rate optimization via waterfilling}
\label{subsec:perf_waterf}
From section \ref{subsec:best_alpha}, the optimal amount of transmitted data energy is derived thanks to eq.(\ref{eq:optimal_alpha_decod_0}), (\ref{eq:optimal_alpha_decod_1}), and (\ref{eq:optimal_alpha_decod_5}). It leads to the coefficient $\alpha_{\text{opt}}^G$ that maximizes the ergodic SR of the communication depending on $\mat{G}$. It is a unique power coefficient weighting the $Q$ components of the useful transmitted data.  Since the channel capacity is proportionnal to the subcarrier energy, and since Alice has access only to the instantaneous channel capacity at Bob, she can tune the amount of transmitted data energy at each subcarrier, i.e., she can apply a different weight at each subcarrier, to enhance the instantaneous capacity at Bob. In doing so, at each channel realization, she determines a new set of coefficients, denoted $\boldsymbol\alpha_{\text{w}}^G = [\alpha_{\text{w},0}^G,...,\alpha_{\text{w},Q-1}^G]^T $, that enhances the instantaneous capacity at Bob. Because Bob and Eve channels are independent, enhancing the channel capacity at Bob does not change the ergodic capacity at Eve. This power allocation strategy is described below. 

If $\mat{\alpha}_w^G = [\alpha_{w,0}^G,...,\alpha_{w,Q-1}^G]^T$ is the variable to optimize, the objective function to maximize is:
\begin{equation}
	\operatorname*{arg\,max}_{\alpha_w^G}  f(\mat{\alpha}_w^G) = \left| \spread^H \HB^* \sqrt{\mat{\alpha}_w^G} \HB \spread \right|^2
	\label{eq:objective_fct}
\end{equation}
Eq.(\ref{eq:objective_fct}) corresponds to the numerator of Bob's SINR. The constraints are:
\begin{equation}
0 \leq \alpha_{w,i} ^G\leq 1 , \; \forall \; i = 0,...,Q-1  \label{eq:constraint1}
\end{equation}
\begin{itemize}
	\item The received AN still lies in Bob's null space after optimization
\end{itemize}
\begin{equation}
\left| \spread^H \HB^* \sqrt{\mat{1}- \mat{\alpha}_w^G} \w  \right|^2 -  \left| \spread^H \HB^* \sqrt{\mat{1}- \alpha_{\text{opt}}^G} \w  \right|^2 \leq \epsilon \label{eq:constraint2}
\end{equation}
\begin{itemize}
	\item The total transmitted energy remains unchanged after optimization
\end{itemize}
\begin{equation}
\module{\sqrt{\mat{\alpha}_w^G} \HB^* \spread \mat{x} + \sqrt{\mat{1}-\mat{\alpha}_w^G}\w }^2   - \module{\sqrt{\alpha_{\text{opt}}^G}  \HB^* \spread \mat{x} + \sqrt{1-\alpha_{\text{opt}}^G} \w }^2 \leq \epsilon \label{eq:constraint3}
\end{equation}
\begin{itemize}
	\item The transmitted AN energy remains unchanged after optimization
\end{itemize}
\begin{equation}
	\module{\sqrt{\mat{1}- \mat{\alpha}_w^G}\w}^2 - \module{\sqrt{\mat{1}- \alpha_{\text{opt}}^G}\w}^2 \leq \epsilon \label{eq:constraint4}
\end{equation}
where $\epsilon = 1e^{-6}$ is the constraint tolerance. A maximum of 2500 iterations is allowed to carry out the optimization, with a step tolerance of $1e^{-6}$. The optimization is performed thanks to the interior-point algorithm. \\
The initial vector, depending on the investigated scenario $\mat{G}$, is:
\begin{equation}
	\mat{\alpha}_w^{G,0} =\left[\alpha_{\text{opt}}^G , ... ,\alpha_{\text{opt}}^G \right]^T  \; \in \; \mathbb{R}^{Q\times 1}
\end{equation}

It is worth nothing that this approach is computationally expensive since new weights have to be determined at each channel realization because of the FF environment.

\subsection{Performance summary}
Table \ref{tab-perf-summary} summarizes the main characterstics and performances for the three investigated decoding structures at Eve:
\begin{table}[!thb]
	\caption{Performance summary for the three investigated models}
	\begin{tabular}{*{1}{|>{\raggedright}p{0.55in}}*{3}{|>{\raggedright}p{0.78in}}|}
		\hline
		& \textbf{SDS Decoder} & \textbf{MF Decoder}  & \textbf{OC Decoder} 
		\tabularnewline  \hline
		\textbf{Eve's knowledge} & Despreading matrix: $\spread^H$   & Despreading matrix: $\spread^H$  \\ Equivalent channel: $\HB^*\HE$ &  Despreading matrix: $\spread^H$ \\ Own channel: $\HE$
	 	\tabularnewline \hline
		\textbf{Optimal decoding structure at Eve} &  $\textbf{G} = \spread^H$ & $\textbf{G} = \spread^H \HB \HE^*$ & $\textbf{G} = \spread^H \HE^*$ 
		\tabularnewline \hline
		\textbf{SR expression} & Eq.(\ref{eq:SR_anal2_decod_0})  &  Eq.(\ref{eq:SR_anal2_decod_1}) & Eq.(\ref{eq:SR_anal2_decod_5})
		\tabularnewline \hline
		\textbf{Performance} & Highest SR values since very poor decoding performance at Eve. &  Lowest SR values since matched filtering at Eve, leading to a frequency diversity gain. SINR about U times bigger compared to the two other models.&  Very similar performances than for the SDS decoder. However, slightly lower SR values for high AWGN energy at Eve, and high $\alpha$. Exact same SNR required at Bob to guarantee a desired SR than for the SDS decoder.
		\tabularnewline \hline
	\end{tabular}
	\label{tab-perf-summary}
\end{table}

\section{Simulation Results} \label{sec:simulation-results}
In this section, simulation results obtained with \textsc{Matlab} are presented. A bit stream is QAM-modulated and the AN signal is generated. The transmitted signal goes through Bob and Eve Rayleigh-fading channels. At the receiver, the SINRs are computed in order to obtain the capacities and thus, the secrecy rates. A Monte Carlo simulation is conducted with 1000 realizations. At each iteration, the channel is updated (i.e., FF assumption) and the SR is calculated. The ergodic SR is obtained by averaging over these 1000 realizations. 
\subsection{Model performances}
\begin{figure}[h!t]
	\centering
	\includegraphics[width=1\linewidth]{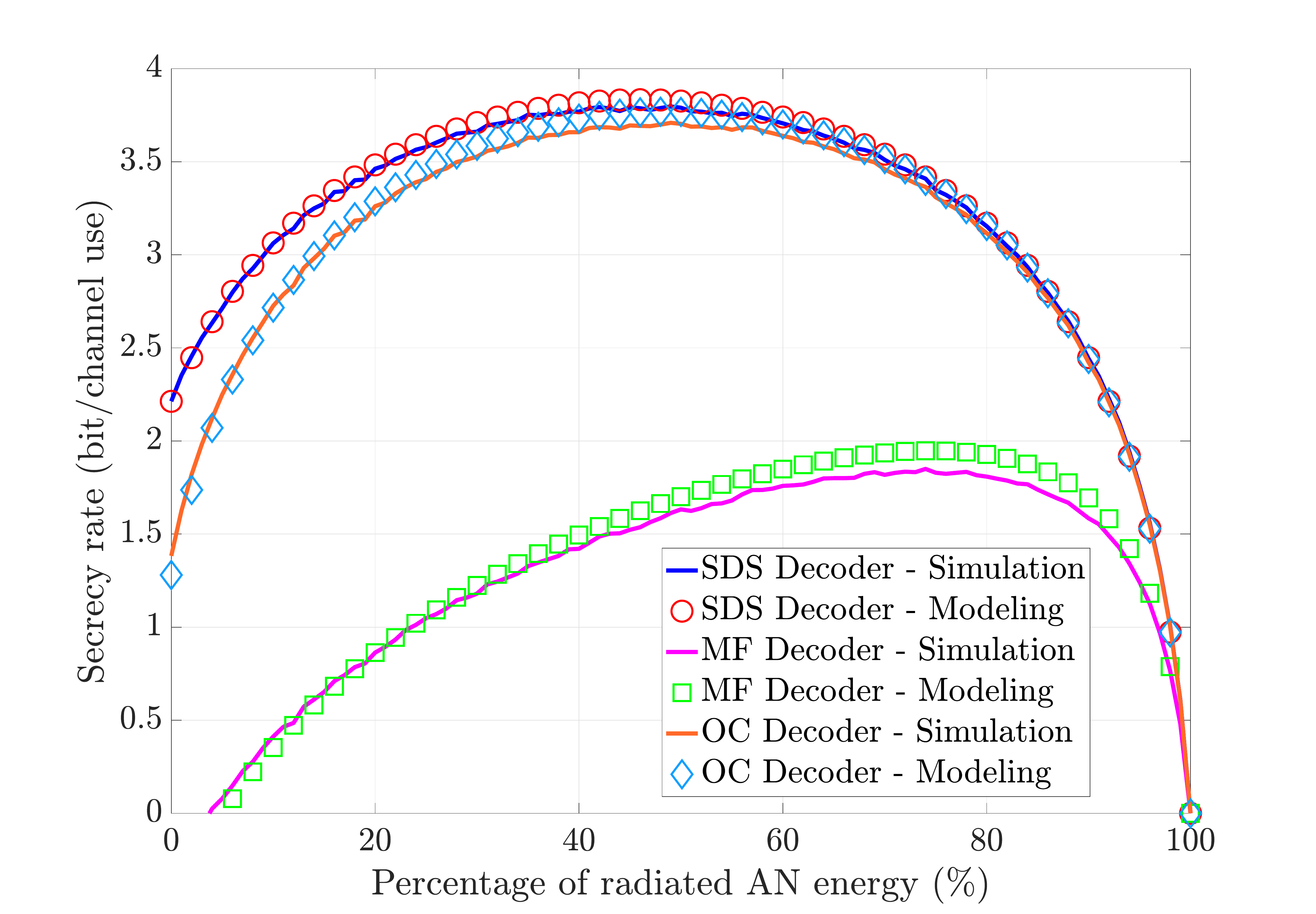}
	\caption{Models vs simulations, $\delta_B = 10$dB,  $\delta_E = 10$dB, BOR = 4}
	\label{fig_modelvssimu}
\end{figure}
Fig.\ref{fig_modelvssimu} shows the secrecy performances as a function of $1-\alpha$, i.e., the AN energy injected, for the three investigated scenarios: Eve implementing a simple despreading (SDS Decoder), Eve implementing a matched filter (MF Decoder), and Eve knowing her own channel (OC Decoder). Fig.\ref{fig_modelvssimu} also outlines a comparaison between the simulation curves (lines) and the analytic ones (markers).

First, it can be seen that the analytical models given by (\ref{eq:SR_anal2_decod_0}), (\ref{eq:SR_anal2_decod_1}), and (\ref{eq:SR_anal2_decod_5}) well approximate the simulation curves and remain tight upper bounds for all scenarios.  In addition, one can notice the importance of the AN addition on the SR. In fact, one can observe a SR enhancement with the addition of AN except for very high percentages of AN sent, i.e., when $1-\alpha \to 1$, or for very low percentages of AN sent, i.e., $1-\alpha \to 0$. Furthermore, for all three models, SR$\to 0$ when $1-\alpha \to 1$ since the SINR at Bob and Eve drops to zero. As anticipated from sections \ref{sec:same-decoding-strucure-as-bob} and \ref{sec:own-channel-knowledge}, high SR values are obtained, i.e., low decoding performance at Eve, when she has the same capabilities as Bob, and when she only knows her own channel. It is also observed that these two scenarios exhibit very similar behaviours except when $1-\alpha \to 0$, as explained in section \ref{sec:own-channel-knowledge}. Finally, one can observe lower SR values when Eve implements a matched filtering decoding structure. This can be understood from (\ref{eq:rx_eve_filt1}) where it is  noticed that each transmitted data symbol is afffected by a real gain at Eve such that it benefits from a frequency diversity gain, leading to higher decoding performances at Eve, and so, lower SR values. In fact, Eve SINR is about U times larger with the MF decoder compared to the SDS and the OC decoders.

\begin{figure}[h!t]
	\centering
	\includegraphics[width=1\linewidth]{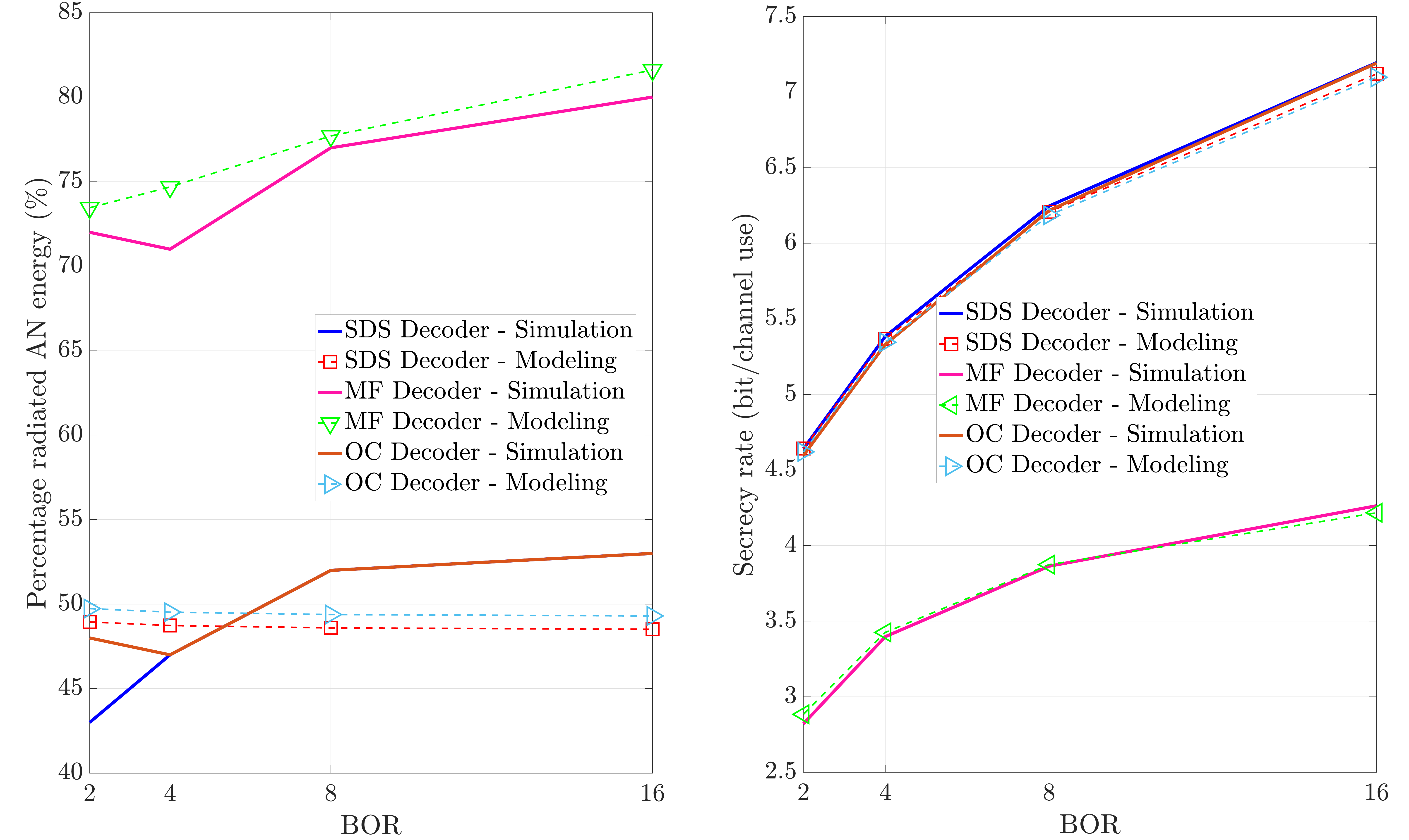}
	\caption{Optimal of AN energy to inject, $\delta_B = 15$dB,  $\delta_E = 15$dB}
	\label{fig_optimalAN}
\end{figure}
The left part of Fig.\ref{fig_optimalAN} illustrates the values of $\alpha_{\text{opt}}$ given by (\ref{eq:optimal_alpha_decod_0}), (\ref{eq:optimal_alpha_decod_1}), and (\ref{eq:optimal_alpha_decod_5}) that maximize the  ergodic SR determined from the closed-form approximations (\ref{eq:SR_anal2_decod_0}), (\ref{eq:SR_anal2_decod_1}), and (\ref{eq:SR_anal2_decod_5}), as well as obtained from the numerical simulations, as a function of the BOR. There is a slight discrepency between the analytical estimations of the optimal amount of data energy to inject using (\ref{eq:optimal_alpha_decod_0}), (\ref{eq:optimal_alpha_decod_1}), and (\ref{eq:optimal_alpha_decod_5}), and its numerical estimation. However, the resulting analytical SR does not differ much from the maximal SR obtained in simulation, as it can be observed on the right part of Fig.\ref{fig_optimalAN}. Indeed, as observed in Fig.\ref{fig_modelvssimu}, the SR is a function that varies slowly about its maximum, for all models. So, for a given BOR value, Alice can make a rough determination of $\alpha_{\text{opt}}^G$ depending on Eve decoding structure, and therefore the available SR, if $\delta_B$ and $\delta_E$ are known. One can also note that much more AN power should be injected to maximize the SR when Eve matched filters the received signal compared to the two other scenarios.

\begin{figure}[h!t]
	\centering
	\captionsetup{justification=centering}
	\includegraphics[width=1\linewidth]{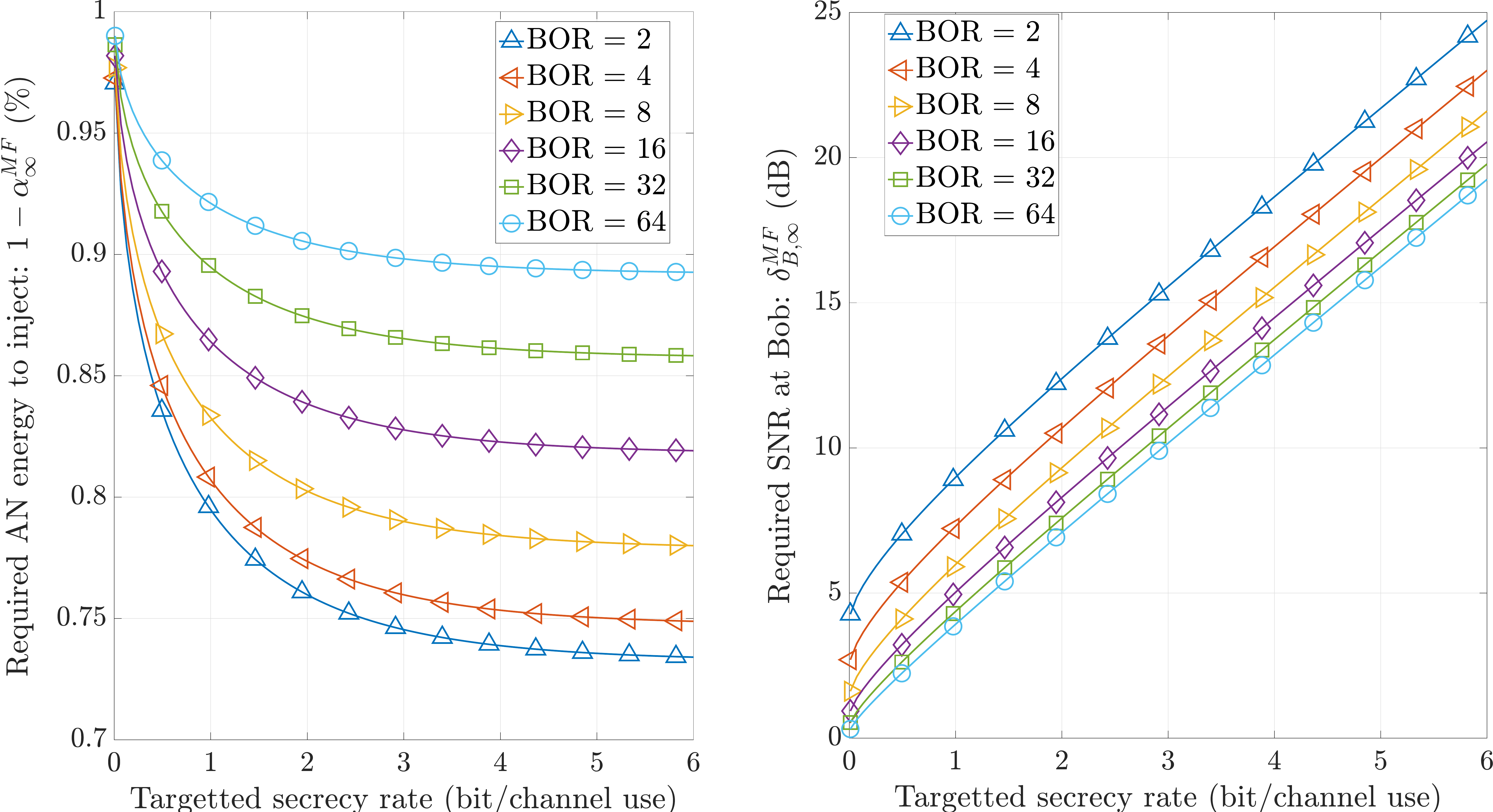}
	\caption{Guaranteeing SR, $\delta_E = \infty$}
	\label{fig_targettedSR}
\end{figure} 
Fig.\ref{fig_targettedSR} illustrates the discussion in section \ref{sec:required-snr-at-bob-for-a-targetted-sr}, for the scenario where Eve implements a matched filter. Eve's SNR is set to $\delta_E = \infty$ such that it represents the conditions under which a desired SR is guaranteed with the proposed FD TR SISO OFDM precoder scheme.

The left part of Fig.\ref{fig_targettedSR} shows the required amount of AN energy to inject (i.e., $1-\alpha_{\infty}^{MF}$) for different BOR values, as a function of the targetted the SR (i.e., $\Delta$), resulting from eq.(\ref{eq:alpha_infty_MF}). One can observe that less AN energy has to be injected when the BOR value decreases, for a fixed targetted SR. It is also worth to note that, when the targetted SR increases, the amount of AN to inject decreases. 

The right part of Fig.\ref{fig_targettedSR} represents the required SNR values at Bob, i.e., $\gamma_{B,\infty}^{MF}$ from eq.(\ref{eq:snr_calibration_bob_MF_infinite}), when $\alpha$ is replaced by its expression (\ref{eq:alpha_infty_MF}). It is observed that a positive SR can always be garantueed, even for moderate SNR at Bob. In addition, one can observe that lower SNR values are required at Bob for higher BOR values since higher TR gains are obtained when the BOR increases. One can also see that the required SNR linearly increases with an increase of the targetted SR.

\subsection{Waterfilling optimization performances}
\begin{figure}[h!t]
	\centering
	\includegraphics[width=1\linewidth]{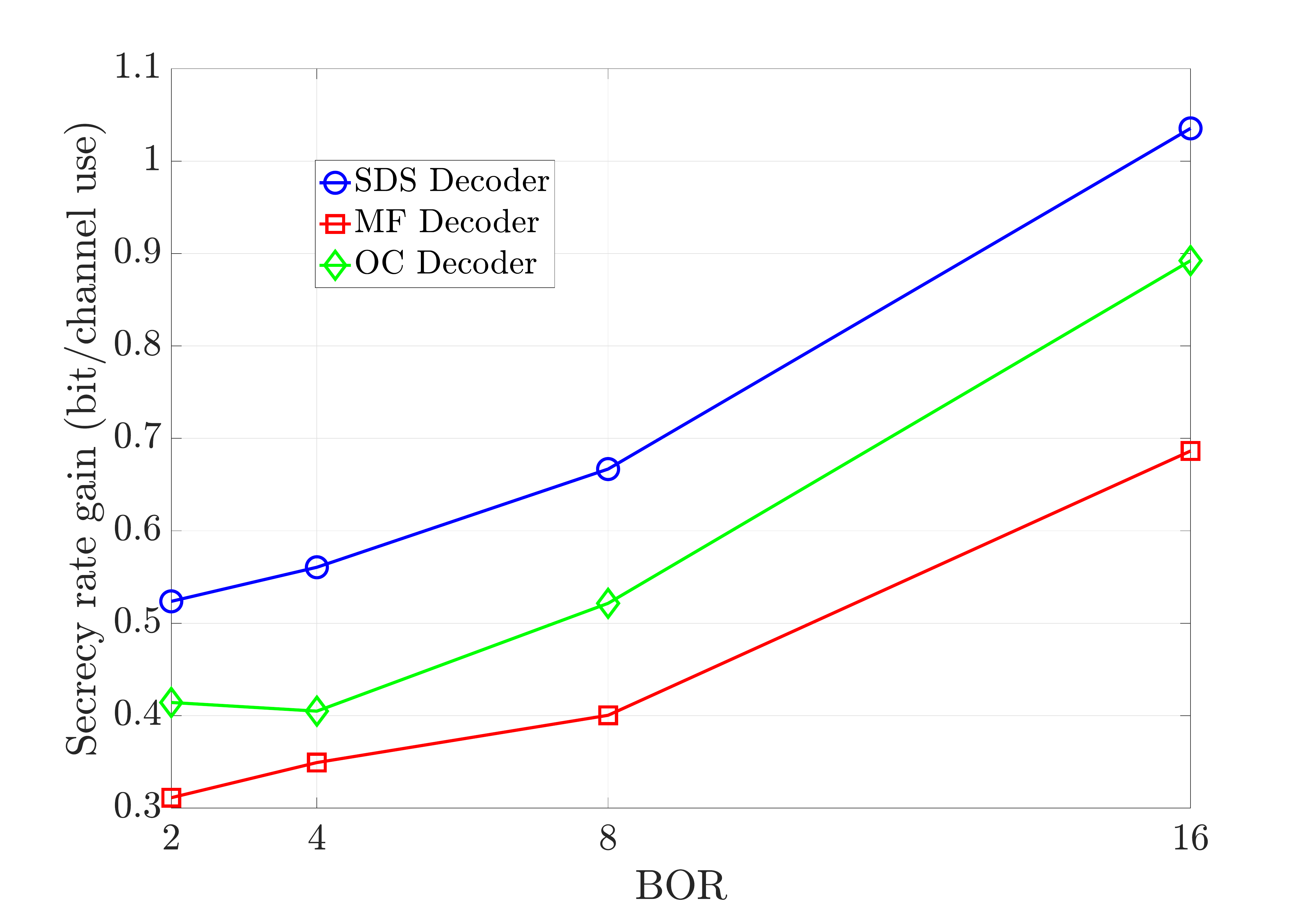}
	\caption{SR gain thanks to waterfilling optimization, $\delta_B  = 15$dB, $\delta_E = 15$dB}
	\label{fig_waterfilling_opt}
\end{figure}
Fig. \ref{fig_waterfilling_opt} presents the SR enhancement thanks to the waterfilling optimization. The SR gain is defined as the difference between the maximal SR obtained after and before optimization. As a reminder, before and after optimization, the mean energy radiated dedicated to the useful data remains unchanged, and the AN signal always remains in Bob's null space. The optimal amount of data energy to inject is computed thanks to (\ref{eq:optimal_alpha_decod_0}), (\ref{eq:optimal_alpha_decod_1}), and (\ref{eq:optimal_alpha_decod_5}) in order to ensure a maximal ergodic SR. The SR is then further increased via the waterfilling optimization procedure, as described in section \ref{subsec:perf_waterf}. As it can be observed, there is an increase of the SR gain for all three models and all BOR values thanks to the waterfilling. 

\section{Conclusions}\label{sec:conclusions}
In this paper, a new scheme is introduced in order to establish a secure communication at the physical layer between a base station, Alice, and a legitimate user, Bob, in the presence of a passive eavesdropper, Eve. Alice uses a time reversal precoder, implemented in the frequency domain with OFDM, to add to the transmitted data an artificial noise that lies in the null-space of Bob but degrade Eve’s channel. The proposed technique only requires a single transmit antenna and is therefore well suited for devices with limited capabilities, such as in IoT for instance.

The ergodic secrecy rate performance is analytically derived, assuming Rayleigh channels, for three different optimal decoding structures at Eve, whose implementation depends on the amount of CSI she can estimate, which in turn depends on the handshake procedure of the considered protocol. The obtained analytical formulations allow Alice to determine the optimal amount of artificial noise energy to inject in order to maximize the secrecy rate. The performance depends on the communication parameters but can be tuned thanks to the back-off rate factor (i.e., sampling rate to symbol rate ratio), used while implementing the time reversal precoder.

Under the assumptions of fast-fading and uncorrelated channels, it is shown that a positive secrecy rate can be guaranteed even when Eve’s SNR is infinite, for moderate values of Bob’s SNR. For instance, with an upsampling of 8, a secrecy rate of 0.75 and 2.2 bits/channel use is obtained with a Bob’s SNR of 5 dB and 10 dB, respectively, with Eve’s SNR is infinite. Furthermore, Alice can be aware of this guaranteed secrecy rate if she knows Bob’s SNR. She can thus communicate while not exceeding this secrecy rate and therefore ensures the secrecy of the communication. Finally, an enhancement of this scheme is proposed via an optimal power allocation strategy over the subcarriers depending on the instantaneous CSI.

This paper shows, consequently, with analytical and simulation results, that a scheme exploiting only frequency degrees of freedom can achieve a positive ergodic secrecy rate to considerably jeopardize any attempt of an eavesdropper to retrieve the data. This approach can be easily integrated into existing standards based on OFDM and does not necessitate extra hardware. However, a perspective of this work is to extend it to multiple antenna systems to assess the benefit of the extra spatial degree of freedom.

\appendices
\section{SINR derivation at Bob}\label{sec:sinr-derivation-app}

\subsection{Data term}\label{sec:data-term-app}
\begin{equation}
	\begin{split}
		\EX{|B_1|^2} =& \EX{\module{\sqrt{\alpha} \; \spread^H \module{\HB}^2 \spread}^2} \\
		\EX{|B_{1,n}|^2} =&\EX{\left|\frac{\sqrt{\alpha}}{U}\sum_{i=0}^{U-1} \left| h_{\text{B}, n + iN}\right|^2\right|^2}  \\
		=& \frac{\alpha}{U^2} \EX{\left(\sum_{i=0}^{U-1} \left| h_{\text{B}, n + iN}\right|^2\right) \left(\sum_{j=0}^{U-1} \left| h_{\text{B}, n + jN}\right|^2\right)^H}\\
		=&  \frac{\alpha}{U^2} \Big(\EX{\sum_{i=0}^{U-1}\left| h_{\text{B}, n + iN}\right|^4} +\\
		& \EX{\sum_{i=0}^{U-1}\left| h_{\text{B}, n + iN}\right|^2}\EX{\sum_{\substack{j=0 \\ j\neq i}}^{U-1} | h^*_{\text{B}, n + jN}|^2} \Big) \\
		=& \frac{\alpha}{U^2} \left( 2U + U(U-1)\right) = \frac{\alpha (U+1)}{U}
	\end{split}
	\label{eq:appA:data_bob-app}
\end{equation}
where we used the fact that $\EX{\left| h_{\text{B}, n + iN}\right|^2} = 1$ and $\EX{\left| h_{\text{B}, n + iN}\right|^4} = 2$ since $\HB \sim \mathcal{CN}(0,1)$.

\subsection{AWGN term}\label{sec:awgn-term-app}
\begin{equation}
	\begin{split}
		\EX{|B_2|^2} &=  \EX{\module{\spread^H \vb}^2} \\
		&= \EX{\left(\spread^H \vb \right)\left(\spread^H \vb \right)^H} \\
		&=\EX{\spread^H \vb \vb^* \spread } \\
		\EX{|B_{2,n}|^2} &= \frac{1}{U} \EX{\sum_{i=0}^{U-1} |v_{\text{B}, n + iN}|^2} = \sigma^2_{\text{V,B}}
	\end{split}
	\label{eq:appA:noise_bob-app}
\end{equation}

\section{SINR derivation at Eve}\label{sec:at-the-unintended-position-app}
\subsection{SDS Decoder}\label{sec:same-decoding-structure-as-bob-app}
\subsubsection{Data term}\label{sec:data-term-app-1}
\begin{equation}
	\begin{split}
		\EX{|\textbf{E}_{1}^{SDS}|^2} &= \EX{\module{\sqrt{\alpha}\spread^H \HE\HB^* \spread}^2} \\
		\EX{|E_{1,n}^{SDS}|^2}&=\alpha \EX{\frac{1}{U^2} \sum_{i=0}^{U-1} \left| h_{\text{E}, n + iN} \right|^2 \left| h^*_{\text{B}, n + iN}\right|^2 } \\
		&= \frac{\alpha}{U}
	\end{split}
	\label{eq:appA:data_eve_filt0-app}
\end{equation}

\subsubsection{AWGN term}\label{sec:awgn-term-app-1}
\begin{equation}
	\begin{split}
		\EX{|\textbf{E}_{2}^{SDS}|^2} &=  \EX{\module{\spread^H \ve}^2} \\
		&=\EX{\spread^H \ve \ve^* \spread } \\
		\EX{|E_{2,n}^{SDS}|^2} &= \frac{1}{U} \EX{\sum_{i=0}^{U-1} |v_{\text{E}, n + iN}|^2} = \sigma^2_{\text{V,E}}
	\end{split}
	\label{eq:appA:noise_eve_filt0-app}
\end{equation}

\subsubsection{AN term}\label{sec:an-term-app-1}
\begin{equation}
	\begin{split}
		\EX{|\textbf{E}_{3}^{SDS}|^2} &=  \EX{\module{\sqrt{1-\alpha}\spread^H \HE \w}^2} \\
		&=(1-\alpha)\EX{\spread^H \HE\textbf{H}^*_{\text{E}} \w\w^* \spread } \\
		\EX{|E_{3,n}^{SDS}|^2}  &= \frac{1-\alpha}{U} \EX{\sum_{i=0}^{U-1} |h_{\text{E}, n + iN}w_{n + iN}|^2} = \frac{1-\alpha}{U}
	\end{split}
	\label{eq:appA:an_eve_filt0-app}
\end{equation}

\subsection{MF Decoder}\label{sec:matched-filtering-app}
\subsubsection{Data term}\label{sec:data-term-app-2}
\begin{multline}
	\EX{|E_{1,n}^{MF}|^2} = \alpha \EX{\left|\frac{1}{U}\sum_{i=0}^{U-1} \left| h_{\text{B}, n + iN}\right|^2 \left| h_{\text{E}, n + iN}\right|^2\right|^2} \\
	=\frac{\alpha}{U^2} \mathbb{E} \Bigg[\sum_{i=0}^{U-1} \left| h_{\text{B}, n + iN}\right|^4 \left| h_{\text{E}, n + iN}\right|^4 \\
	+ \sum_{i=0}^{U-1}\sum_{\substack{j=0 \\ j\neq i}}^{U-1}  \left| h_{\text{B}, n + jN}\right|^2 \left| h_{\text{E}, n + iN}\right|^2 \left| h^*_{\text{B}, n + iN}\right|^2 \left| h^*_{\text{E}, n + jN}\right|^2 \Bigg] \\
	= \frac{\alpha}{U^2} \left(U.2.2 + U(U-1) \right) = \frac{\alpha (U+3)}{U}
	\label{eq:data_eve_filt1-app}
\end{multline}
where we used the fact that $\EX{\left| h_{\text{E}, n + iN}\right|^2} = 1$ and $\EX{\left| h_{\text{E}, n + iN}\right|^4} = 2$ since $\HE \sim \mathcal{CN}(0,1)$.

\subsubsection{AWGN term}\label{sec:awgn-term-app-2}
\begin{equation}
	\begin{split}
		\EX{|\textbf{E}_{2}^{MF}|^2} &=  \EX{\module{\spread^H \HE^* \HB \ve}^2} \\
		&=\EX{\spread^H   \HE \HE^* \HB\HB^*  \ve \ve^* \spread } \\
		\EX{|E_{2,n}^{MF}|^2} &= \frac{1}{U} \EX{\sum_{i=0}^{U-1} |h_{\text{E}, n + iN}|^2 |h_{\text{B}, n + iN}|^2 |v_{\text{E}, n + iN}|^2} = \sigma^2_{\text{V,E}}
	\end{split}
	\label{eq:noise_eve_filt1-app}
\end{equation}

\subsubsection{AN term}\label{sec:an-term-app-2}
The component $\textbf{A}_{3,n}$ depends on $\textbf{w}$ and $\HB$ which are correlated via the AN design (\ref{eq:an_cond}). The expectation is therefore not straightforward to compute. Remembering that $\mat{A} = \spread^H \HB$. Omitting the $1-\alpha$ as well as the normalization  factor in eq.(\ref{eq:an_w}), the AN term at Eve is given by:
\begin{equation}
	\begin{split}
	\vect{v}&=\mat{A} |\mat{H}_E|^2 \vect{w} \label{eq:an_decod1_a}\\
	&=\mat{U} \mat{\Sigma}\mat{V}_1^H |\mat{H}_E|^2 \mat{V}_2 \vect{w}'
	\end{split}
\end{equation}
Note that $\vect{w}'$ is independent of the other random variables and has a unit covariance matrix. Therefore, it can be shown that:
\begin{align}
\mathbb{E}\left(\vect{v}\vect{v}^H\right)&=\mathbb{E}\left(\mat{U} \mat{\Sigma}\mat{V}_1^H |\mat{H}_E|^2 \mat{V}_2 \mat{V}_2^H|\mat{H}_E|^2\mat{V}_1 \mat{\Sigma}^H   \mat{U}^H\right)
\end{align}
Let's rewrite $|\HE|^2=\sum_{q=1}^Q|H_{E,q}|^2 \vect{e}_q \vect{e}_q^T $ where $\vect{e}_q$ is an all zero vector except a $1$ at row $q$:
\begin{equation}
	\begin{split}
	\mathbb{E}\left(\vect{v}\vect{v}^H\right)=&\sum_{q=1}^Q\sum_{q'=1}^Q\mathbb{E}(|H_{E,q}|^2|H_{E,q'}|^2)\\
	&\mathbb{E}\left(\mat{U} \mat{\Sigma}\mat{V}_1^H  \vect{e}_q \vect{e}_q^T \mat{V}_2 \mat{V}_2^H \vect{e}_{q'} \vect{e}_{q'}^T\mat{V}_1 \mat{\Sigma}^H   \mat{U}^H\right)\\
	=&\sum_{q=1}^Q\mathbb{E}\left(\mat{U} \mat{\Sigma}\mat{V}_1^H  \vect{e}_q \vect{e}_q^T \mat{V}_2 \mat{V}_2^H \vect{e}_{q} \vect{e}_{q}^T\mat{V}_1 \mat{\Sigma}^H   \mat{U}^H\right)\\
	&+\mathbb{E}\left(\mat{U} \mat{\Sigma}\mat{V}_1^H  \mat{V}_2 \mat{V}_2^H \mat{V}_1 \mat{\Sigma}^H   \mat{U}^H\right)
	\end{split}
\end{equation}
where the second term cancels out since $\mat{V}_2^H \mat{V}_1=\mat{0}$. Since all elements of $\vect{v}$ have same variance, the following holds:
\begin{equation}
	\begin{split}
	\frac{1}{N}\mathbb{E}\left(\|\vect{v}\|^2\right)&=\frac{1}{N}\mathbb{E}\left(\vect{v}\vect{v}^H\right) \\
	&=\frac{1}{N}\mathbb{E} \left( \mat{\Sigma}^2\mat{V}_1^H  \sum_{q=1}^Q\left(\vect{e}_q \vect{e}_q^T \mat{V}_2 \mat{V}_2^H \vect{e}_{q} \vect{e}_{q}^T\right)\mat{V}_1 \right)
	\end{split}
\end{equation}
Let's rewrite $\mat{V}_1=\sum_{l}\vect{e}_{l}\vect{v}_{1,l}^H$ where $\vect{v}_{1,l}^H$ is the $l$-th row of $\mat{V}_1$ (of dimension $N\times 1$) with only one nonzero element.
\begin{equation}
	\begin{split}
	\frac{1}{N}\mathbb{E}\left(\|\vect{v}\|^2\right)&=\frac{1}{N}\sum_{q=1}^Q\sum_{l}\sum_{l'}\mathbb{E} \left( \mat{\Sigma}^2\vect{v}_{1,l}\vect{e}_{l'}^T  \vect{e}_q \vect{e}_q^T \mat{V}_2 \mat{V}_2^H \vect{e}_{q} \vect{e}_{q}^T\vect{e}_{l}\vect{v}_{1,l}^H \right)\\
	&=\frac{1}{N}\sum_{q=1}^Q \mathbb{E} \left( \mat{\Sigma}^2\vect{v}_{1,q}\vect{e}_q^T \mat{V}_2 \mat{V}_2^H \vect{e}_{q} \vect{v}_{1,q}^H \right)
	\end{split}
\end{equation}
Let's rewrite $\mat{V}_2=\sum_{l}\vect{e}_{l}\vect{v}_{2,l}^H$ where $\vect{v}_{2,l}^H$ is the $l$-th row of $\mat{V}_2$ (of dimension $Q-N\times 1$) with $U-1$ nonzero elements:
\begin{equation}
	\begin{split}
	\frac{1}{N}\mathbb{E}\left(\|\vect{v}\|^2\right)&=\frac{1}{N}\sum_{q=1}^Q\sum_l\sum_{l'} \mathbb{E} \left( \mat{\Sigma}^2\vect{v}_{1,q}\vect{e}_q^T \vect{e}_{l}\vect{v}_{2,l}^H \vect{v}_{2,l'} \vect{e}_{l'}^T \vect{e}_{q} \vect{v}_{1,q}^H \right)\\
	&=\frac{1}{N}\sum_{q=1}^Q \mathbb{E} \left( \| \vect{v}_{2,q}\|^2\vect{v}_{1,q}^H\mat{\Sigma}^2\vect{v}_{1,q}  \right)
	\end{split}
\end{equation}

where $\vect{v}_{1,q}^H \mat{\Sigma}^2 \vect{v}_{1,q} \coloneqq  \| \vect{v}_{1,q}\|^2 \sigma_n^2$ is a scalar. Therefore:

\begin{align}
\frac{1}{N}\mathbb{E}\left(\|\vect{v}\|^2\right) &=\frac{1}{N}\sum_{q=1}^Q \mathbb{E} \left( \| \vect{v}_{2,q}\|^2 \| \vect{v}_{1,q}\|^2 \sigma_n^2  \right)
\end{align}
Since $\mat{V}$ forms an orthonormal basis, i.e., $\mat{V}^H \mat{V} = \mat{I}_Q$, it is found that $ \| \vect{v}_{1,q}\|^2 +  \| \vect{v}_{2,q}\|^2 = 1$. Then:
\begin{align}
\frac{1}{N}\mathbb{E}\left(\|\vect{v}\|^2\right) &=\frac{1}{N}\sum_{q=1}^Q \mathbb{E} \left[ \left( \| \vect{v}_{1,q}\|^2 - \| \vect{v}_{1,q}\|^4 \right) \sigma_n^2  \right]
\label{eq:v_1}
\end{align}
To determine (\ref{eq:v_1}), the transformations performed by the SVD on $\mat{A}$ in order to obtain $\vect{v}_{1,q}$ and $\sigma^2_n$ need to be determined. One can show that:
\begin{equation}
	\sigma_n = \sqrt{\sum_{i=1}^{U} \left| z_{(n-1)U+i}\right|^2} \; , n = 1...N
\end{equation}
where $z_i  = z_{i,x} + jz_{i,y} \sim \mathcal{CN}(0,\frac{1}{U})$. Therefore:
\begin{equation}
	\mathbb{E} \left[\sigma_n^2 \right] = 1
\end{equation}
Without loss of generality, $\mathbb{E} \left[ \| v_1\|^2 \right]$ and $\mathbb{E} \left[ \| v_1\|^4 \right]$  can be computed since all components of $\mat{V}_1$ are identically distributed:
\begin{equation}
	\begin{split}
	\mathbb{E} \left[ \| v_1\|^2 \right] &= \mathbb{E}\left[ \left| \frac{z_1^*}{\sigma_1}\right|^2\right]  = \frac{1}{U}
	\end{split}
\end{equation}
For the moment of order 4, knowing that $\mathbb{E}\left[ \left| z_i \right|^4\right] = \frac{2}{U^2}$:
\begin{equation}
	\begin{split}
	\mathbb{E} \left[ \| v_1\|^4 \right] &= \mathbb{E}\left[ \left| \frac{z_1^*}{\sigma_1}\right|^4\right] \\
	&=  \mathbb{E} \left[   \frac{\left| z_1 \right|^4 }{  \left( \sum_{i=1}^{U} \left| z_i\right|^2 \right)^2 }  \right] \label{eq:moment_4_1} \\
	&= \mathbb{E} \left[   \frac{\left| z_1 \right|^4 }{  \sum_{i=1}^{U} \left| z_i\right|^4 + 2 \sum_{i=1}^{U} \sum_{j<i} \left|z_i\right|^2 \left|z_j\right|^2 }  \right]\\
	&= \frac{2}{U(U+1)}
	\end{split}
\end{equation}
Finally, eq.(\ref{eq:v_1}) can be computed as:
\begin{equation}
	\frac{1}{N}\mathbb{E}\left(\|\vect{v}\|^2\right)=\frac{1}{N} \sum_{q=1}^{Q} \left[ \left( \frac{1}{U} - \frac{2}{U(U+1)} \right) 1\right] = \frac{U-1}{U+1} \label{eq:final_an_result_no_correction}
\end{equation}
Keeping into account the normalization factors, it follows:
\begin{equation}
	\EX{|E_{3,n}^{MF}|^2} = (1-\alpha)\frac{1}{U-1}\frac{U-1}{U+1}  =  \frac{1-\alpha}{U+1}
	\label{eq:an_eve_filt1-app}
\end{equation}

\subsection{OC Decoder}\label{sec:own-channel-knowledge-app}
\subsubsection{Data term}\label{sec:data-term-app-3}
\begin{multline}
	\EX{|E_{1,n}^{OC}|^2} = \alpha \EX{\left|\frac{1}{U}\sum_{i=0}^{U-1} h_{\text{B}, n + iN}^* \left| h_{\text{E}, n + iN}\right|^2\right|^2} \\
	=\frac{\alpha}{U^2} \mathbb{E}\Bigg[ \sum_{i=0}^{U-1} \left| h_{\text{B}, n + jN}\right|^2 \left| h_{\text{E}, n + iN}\right|^4 \\
	+ \sum_{i=0}^{U-1}\sum_{\substack{j=0 \\ j\neq i}}^{U-1}  h_{\text{B}, n + jN} h^*_{\text{B}, n + iN} \left| h_{\text{E}, n + jN}\right|^2 \left| h^*_{\text{E}, n + iN}\right|^2 \Bigg] \\
	= \frac{\alpha}{U^2} \left(U.2.1 + U(U-1).1.1.0\right) = \frac{2\alpha}{U}
	\label{eq:data_eve_filt5-app}
\end{multline}

\subsubsection{AWGN term}\label{sec:awgn-term-app-3}
\begin{equation}
	\begin{split}
		\EX{|\textbf{E}_{2}^{OC}|^2} &=  \EX{\module{\spread^H \HE^* \ve}^2} \\
		&=\EX{\spread^H   \left|\HE\right|^2  \left|\ve\right|^2 \spread } \\
		\EX{|E_{2,n}^{OC}|^2} &= \frac{1}{U} \EX{\sum_{i=0}^{U-1} |h_{\text{E}, n + iN}|^2 |v_{\text{E}, n + iN}|^2} = \sigma^2_{\text{V,E}}
	\end{split}
	\label{eq:noise_eve_filt5-app}
\end{equation}

\subsubsection{AN term}\label{sec:an-term-app-3}
\begin{equation}
	\begin{split}
		\EX{|\textbf{E}_{3}^{OC}|^2} &=  \EX{\module{\sqrt{1-\alpha}\spread^H \left|\HE\right|^2 \w}^2} \\
		&=(1-\alpha)\EX{\spread^H \left|\HE\right|^2 \w\w^* \left|\HE^*\right|^2\spread } \\
		\EX{|E_{3,n}^{OC}|^2}  &= \frac{1-\alpha}{U} \EX{\sum_{i=0}^{U-1} |h_{\text{E}, n + iN}|^4 |w_{n + iN}|^2} = \frac{2(1-\alpha)}{U}
	\end{split}
	\label{eq:an_eve_filt5-app}
\end{equation}



\bibliographystyle{IEEEtran}
\bibliography{biblio}




\end{document}